\documentclass[fleqn,usenatbib]{mnras}

\usepackage{graphicx}
\usepackage{dcolumn}
\usepackage{bm}
\usepackage{graphics, caption}
\usepackage[utf8]{inputenc}
\usepackage[T1]{fontenc}
\usepackage{amsmath,booktabs,braket,cleveref,mathptmx,xcolor,siunitx}
\usepackage[version=4]{mhchem} 
 \usepackage[normalem]{ulem}
\usepackage{array}

\newcommand{\abinitio}{\emph{ab initio}}
\newcommand{\Duo}{{\sc Duo}}

\newcommand{\Marvel}{{\sc Marvel}}
\newcommand{\MARVEL}{{\sc Marvel}}

\newcommand{\cm}{cm$^{-1}$}

\def\Xstate{{X~${}^{1}\Sigma_{\rm g}^{+}$}}
\def\Astate{{A~${}^{1}\Pi_{\rm u}$}}
\def\Dstate{{D~${}^{1}\Sigma_{\rm u}^{+}$}}
\def\Cstate{{C~${}^{1}\Pi_{\rm g}$}}
\def\Cprimestate{{C$^\prime$~${}^{1}\Pi_{\rm g}$}}
\def\Bstate{{B~${}^{1}\Delta_{\rm g}$}}
\def\Bprimestate{{B$^\prime$~${}^{1}\Sigma_{\rm g}^{+}$}}
\def\Estate{{E~${}^{1}\Sigma_{\rm g}^{+}$}}
\def\Fstate{{F~${}^{1}\Pi_{\rm u}$}}
\def\Onedeltastate{{1~${}^{1}\Delta_{\rm u}$}}

\def\LLname{\texttt{8states}}

\def\astate{{a~${}^{3}\Pi_{\rm u}$}}
\def\bstate{{b~${}^{3}\Sigma_{\rm g}^{-}$}}
\def\cstate{{c~${}^{3}\Sigma_{\rm u}^{+}$}}
\def\dstate{{d~${}^{3}\Pi_{\rm g}$}}
\def\estate{{e~${}^{3}\Pi_{\rm g}$}}
\def\fstate{{f~${}^{3}\Sigma_{\rm g}^+$}}
\def\gstate{{g~${}^{3}\Delta_{\rm g}$}}

\def\threestate{{3~${}^{3}\Pi_{\rm g}$}}
\def\fourstate{{4~${}^{3}\Pi_{\rm g}$}}

\def\a0{{$a_{\rm 0}$}}

\newcommand{\edited}[1]{#1}

\newcommand{\mc}{\multicolumn}
\newcolumntype{H}{>{\setbox0=\hbox\bgroup}c<{\egroup}@{}}

\newcolumntype{d}{D{.}{.}{-1}}
\newcommand{\noenergy}{7047}
\newcommand{\notrans}{31\,323}
\newcommand{\novalid}{30\,792}
\newcommand{\noelec}{20}
\newcommand{\novibronic}{142}

\renewcommand*\descriptionlabel[1]{\hspace\labelsep\normalfont #1}

\graphicspath{{Figs/}}

\title{An update to the MARVEL dataset and ExoMol line list for $^{12}$C$_2$} 

\author[McKemmish et al.]{
Laura K. McKemmish,$^{1}$\thanks{E-mail: l.mckemmish@unsw.edu.au}
Anna-Maree Syme,$^{1}$
Jasmin Borsovszky,$^{1}$
Sergei N. Yurchenko,$^{2}$\newauthor
Jonathan Tennyson,$^{2}$
Tibor Furtenbacher$^{3}$
and Attila G. Cs\'asz\'ar$^{3}$
\\
$^{1}$School of Chemistry, University of New South Wales, 2052 Sydney\\
$^{2}$Department of Physics and Astronomy, University College London, London WC1E 6BT, United Kingdom\\
$^{3}$Institute  of Chemistry, ELTE E\"otv\"os Lor\'and University and
MTA-ELTE Complex Chemical Systems Research Group,
H-1518 Budapest 112, P.O. Box 32, Hungary
}

\date{Accepted XXX. Received YYY; in original form ZZZ}

\pubyear{2020}

\begin{document}
\label{firstpage}
\pagerange{\pageref{firstpage}--\pageref{lastpage}}
\maketitle

\begin{abstract}

The spectrum of dicarbon (C$_2$) is important in astrophysics
and for spectroscopic studies of plasmas and flames.
The C$_2$ spectrum is characterized by many band systems with new ones still being actively identified;
astronomical observations involve eight of these bands. 
Recently,  Furtenbacher {\it et al.} (2016, Astrophys. J. Suppl., 224, 44) presented a set of 5699 empirical energy levels  for $^{12}$C$_2$,
distributed among 11 electronic states and 98 vibronic bands,
derived from 42 experimental studies and obtained using the \Marvel{} (Measured Active Rotational-Vibrational Energy Levels) procedure. 

Here, we add data from 13 new sources and update data from 5 sources. 
Many of these  data sources characterize high-lying electronic states,
including the newly detected \threestate{} state.
Older studies have been included following improvements in the \Marvel{} procedure which allow their uncertainties to be estimated. These older works in particular determine levels in the \Cstate{} state, the upper state of the insufficiently characterized Deslandres--d'Azambuja (\Cstate{}--\Astate{}) band.  

The new compilation considers a total of \notrans{} transitions and derives \noenergy{} empirical (\Marvel{}) energy levels spanning 20 electronic and 142 vibronic states. 
These new empirical energy levels are used here to update the  \LLname{} C$_2$ ExoMol line list. 
This updated line list is highly suitable for high-resolution cross-correlation studies in astronomical spectroscopy of, for example, exoplanets, as 99.4\% of the transitions with intensities over 
10$^{-18}$ cm\,molecule$^{-1}$ at 1000 K have frequencies determined by empirical energy levels. 

\end{abstract}


\section{Introduction}

The spectroscopy of the dicarbon molecule, C$_2$, has a long history. 
Interestingly, \ce{C2} was originally observed by \citet{1802Wo.C2}, which represents the pre-history of spectroscopy. 
This observation was followed by the identification  \citep{1857Swan.C2}
of the well-known Swan \dstate{} -- \astate{} band system.
The last decade has seen the spectroscopic characterization of 
several new bands of C$_2$, including the first
observation of multiplicity-changing ``intercombination'' bands linking both the
singlet-triplet  \citep{15ChKaBeTa.C2} and triplet-quintet  \citep{11BoSyKnGe.C2} states. 
These observations have allowed the determination of reliable
frequencies of singlet-triplet transitions, which are thought 
to be important in  comets  \citep{00RoHiBu} and are candidates 
for observation in the interstellar medium  \citep{86LeRo}. 
Detection of  triplet-quintet transitions has led to the spectroscopic characterization
of a number of quintet states
for the first time  \citep{11SchmBa,15BoMaGo,17BoViBeKn}. 
In addition, recent experiments detected and characterized a 
number of new triplet bands  \citep{17WeKrNaBa,17KrWeBaNa}. 

\begin{figure}
    \includegraphics[width=0.4\textwidth]{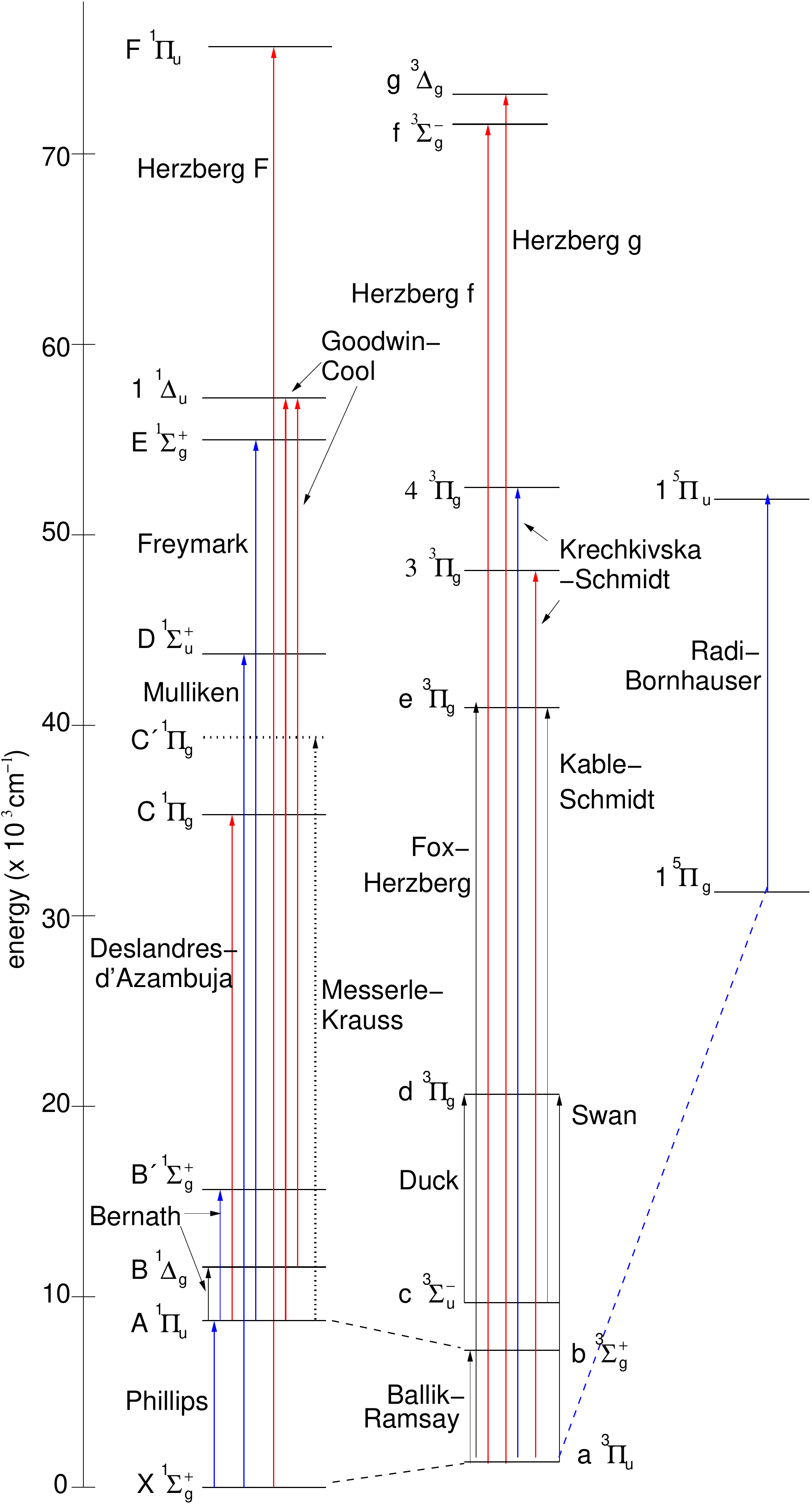}
    \caption{The band system of $^{12}$C$_2$ with its well-established names. 
    The dashed lines represent observed but unnamed intercombination bands; colours indicate newly considered bands (red) and updates (blue).
    The Messerle--Krauss band and the associated \Cprimestate\ state
    are depicted in dots as doubts
    have been raised about their correctness.} 
    \label{fig:PES}
\end{figure}

Astronomically, C$_2$ is unusual in that it has been studied via 
a large number of band systems including the Swan, Phillips, Deslandres--d'Azambuja,
Ballik--Ramsay,  Mulliken and Herzberg-F bands, see \Cref{fig:PES} for the band designations. 
The Swan  \citep{43Swings,89GrVaBl,90LaShDaAr,00RoHiBu} and the
Deslandres--d'Azambuja  \citep{89GrVaBl} band systems 
have been discovered in the spectra of comets when 
models of cometary emission have been found to require no less  than two singlet, 
\Xstate\ and \Astate, and four triplet, \astate, \bstate, \cstate, and 
d~$^{3}\Pi_{\rm g}$, electronic states to explain the observations.
Indeed, two of the intercombination bands mentioned above,
\astate $\rightarrow$ \Xstate\ and 
\cstate $\rightarrow$ \Xstate, are needed to explain the
observed intensities in the Swan band  \citep{00RoHiBu}.

C$_2$ has a strong presence in the solar photosphere where it has
been observed  using the Swan  \citep{05AsGrSaPr},
the Phillips, and the Ballik--Ramsay    \citep{82BrDeGr} bands. 
The Phillips and Ballik--Ramsay bands have also been observed in carbon
stars  \citep{83GoBrCo.C2,90Goorvitc}, while Swan bands 
 have  been observed in peculiar white
dwarfs  \citep{08HaMa.C2,10Kowalski} and the coronae borealis star V coronae
australis  \citep{08RaLa}.

Interstellar C$_2$ has been observed via the infrared Phillips band
 \citep{01GrBlYa.C2,11Iglesias}, while the Swan band emissions can be seen
in the Red Rectangle  \citep{10WeRoLi}. 
Absorption in the Phillips, Mulliken and Herzberg-F bands can be seen in
translucent clouds   \citep{07SoWeThYo}.

These astronomical observations require high-quality laboratory data for their
analysis and interpretation.
Recent spectroscopic studies 
have probed new bands with well-known band systems  \citep{19Nakajima, 18KrWeFrNa}, providing new data on them. 
In addition, recent spectroscopic studies on C$_2$ have used techniques 
yielding improved ionization  \citep{16KrBaWeNa} and dissociation 
energies  \citep{19ViBeBoKn}. 
Theoretical studies also started to provide reliable 
association rates  \citep{19BaSmMc}.
Altogether work on the C$_2$ molecule remains as lively as ever with the 
interpretation of its bonding and spectroscopy remaining somewhat as
a puzzle to conventional chemical physics  \citep{16Macrae}.

\Cref{fig:PES} gives an overview of the observed band systems for $^{12}$C$_2$
with colour used to indicate those explicitly dealt with in this study. 
In response to the needs of astrophysics and other areas of physics, 
\citet{jt736} computed a comprehensive line list for $^{12}$C$_2$ 
as part of the ExoMol project  \citep{jt528}, called the \LLname{} line list.  
This line list was generated by variational solution of the nuclear 
Schr\"odinger equation for the states involved  \citep{jt609} and  
covers the band systems linking the lowest eight electronic states, 
namely the Swan, Phillips, Ballik--Ramsay, Duck, Bernath B and B' bands, 
and the singlet-triplet intercombination lines.
As a precursor to performing these calculations, 
\citet{jt637} performed a \Marvel{} 
(Measured Active Rotational-Vibrational Energy Levels \citep{07CsCzFu.marvel,jt412,12FuCs}, 
see Section~2 for a description) analysis for the $^{12}$C$_2$ isotopologue.
The empirical energies generated by
MARVEL were incorporated in the  $^{12}$C$_2$ \LLname{} line list giving, for example, the most accurate predictions available for the singlet-triplet intercombination lines. 

A number of advances has led us to review and update the $^{12}$C$_2$ \Marvel{} project. 
First, improvements in the \Marvel{} procedure, including
significantly improved error handling  \citep{19ToFuTeCs}, 
was found to influence the results of the original study. 
Second, while the original \Marvel{} study considered 42 sources of 
spectroscopic $^{12}$C$_2$ data, a number of largely older sources  \citep{30DiLo,37FoHe.C2,40HeSu,50Phillips,69HeLaMa,88GoCob,89GoCo} 
were not considered in 2016 as they did not contain
any uncertainty estimates, a requirement for use in the \Marvel{} procedure. 
New combination difference approaches implemented in \Marvel{} allow these 
uncertainties to be accurately estimated. 
These earlier works contain data on states that have not been observed
in more recent studies; in particular
 the studies of 40HeSu  \citep{40HeSu}, 50Phillips  \citep{50Phillips} and 67Messerle  \citep{67Messerle} contain the only published
high resolution $^{12}$C$_2$ spectra of the  Deslandres--d'Azambuja band. 
Finally, and most importantly, a series
of new studies have provided additional data for known bands
\citep{17WeKrNaBa,17KrWeBaNa} and characterized several new bands
for the first time  \citep{11SchmBa,15BoMaGo,17BoViBeKn}.
These sources are combined with those considered previously to produce an updated set of empirical (\Marvel{}) rovibronic energy levels during this study. 
All (new and old) data sources are referenced by band in \Cref{tab:bands} (\emph{vide infra}).


In this work we also present an updated and 
improved version of the $^{12}$C$_2$ \LLname{} ExoMol line list, which incorporates the new and extended \Marvel-derived set of empirical energy levels.

\section{The \Marvel{} procedure}

Details about the \Marvel{} procedure  \citep{jt412,12FuCs,19ToFuTeCs},
built upon the theory of spectroscopic networks (SN)  \citep{11CsFu,16CsFuAr}, 
have been given in recent publications  \citep{14FuArMe.marvel,16CsFuAr,19ToFuTeCs,20ToFuSiCs}.
Therefore, only a brief discussion 
is given here.

The \Marvel{} protocol yields empirical 
rovibronic energies with well-defined provenance and uncertainties; it
starts with the construction of a SN
using the dataset of measured and assigned transitions
collated from the literature. 
Each measured transition must have a \emph{unique} 
(though not necessarily physically relevant) assignment,
which determines its place within the SN, and an uncertainty.
What happens next is basically an inversion of the transitions information,
yielding empirically determined rovibronic
energy levels within each component of the SN.
Along this process validation of the experimental information is performed,
utilizing several elements of network theory.
Recently the algorithms employed by \Marvel{}
have been systematically improved, the relationship of SNs to formal 
network theory considered  \citep{11CsFu,12FuCsxx.marvel,16ArPeFu.marvel}, and the
underlying methodology reviewed  \citep{14FuArMe.marvel,16CsFuAr}.
\Marvel{} has been used to obtain accurate empirical rovibronic energy levels 
with statistically sound uncertainties
for a considerable number of diatomic molecules of astronomical interest  \citep{jt637,jt672,jt722,jt732,jt740,jt764,19DaShJoKa,19FuHoKoSo}. 
These \Marvel{} energy levels are crucial to enabling the generation 
of \Marvel ised line lists (e.g. \cite{jt760}) suitable for high-resolution 
cross-correlation studies of low-signal objects such as exoplanets (e.g. \cite{13BideBr.highres}). 

%

\section{Updated \Marvel{} set of assigned transitions} 

\subsection{Overview}
We have updated the \Marvel{} set of 
assigned transitions for $^{12}$C$_2$ through the inclusion of 
13 new data sources (8 new sources from prior to the original update and 
5 post-2016 sources) and through the revision of transitions 
from six further data sources. 
The number of included transitions has risen from the 2016 values of
23\,251 (22\,937 validated) to \notrans{} (\novalid{} validated). 

We usually added new data to the pre-existing \Marvel{} set of transitions and uncertainties, unless specified otherwise.  The uncertainties used for each source were usually taken from the original paper, 
but increased as required, first for internal self-consistency of the
data within a single data source and then for self-consistency with the 
full \Marvel{} compilation of data.

\begin{table*}
\centering
\caption{\label{tab:marvelinput}
Extract from the 12C-12C\_2020update.marvel.inp input file for \ce{^{12}C2}.}
\footnotesize \tabcolsep=5pt
\begin{tabular}{lllccclcccr}
\\
\toprule
  \mc{1}{c}{1}      &         \mc{1}{c}{2}         &    \mc{1}{c}{3}   &    \mc{1}{c}{4}   &   \mc{1}{c}{5}   &   \mc{1}{c}{6}   &    \mc{1}{c}{7}   &   \mc{1}{c}{8}   &   \mc{1}{c}{9} &   \mc{1}{c}{10} &   \mc{1}{c}{11}  \\
    \midrule
    \multicolumn{1}{c}{$\tilde{\nu}$}  &  \multicolumn{1}{c}{$\Delta\tilde{\nu}$}  &   \mc{1}{c}{State$^\prime$}     &  \multicolumn{1}{c}{$v^\prime$}  &   $J^\prime$ &   $F^\prime$   &     \mc{1}{c}{State$^{\prime\prime}$}    &   \multicolumn{1}{c}{$v^{\prime\prime}$}  &  $J^{\prime\prime}$ &  $F^{\prime\prime}$    &     \mc{1}{c}{ID}  \\
     \midrule
3345.6527 & 0.006132865 & B\^{}1Deltag & 1 & 9 & 1  & A\^{}1Piu & 1 & 10 &  1  & 16ChKaBeTa.167 \\
3347.987 & 0.0093071 & b\^{}3Sigmag- & 4 & 26 & 3 & a\^{}3Piu & 5 & 26 & 3  & 15ChKaBeTa.1594\\
3349.659 & 0.009250366 & B\^{}1Deltag & 1 & 16 & 1 & A\^{}1Piu & 1 & 16 & 1 & 16ChKaBeTa.168\\
3349.8868 & 0.0015 & B\^{}1Deltag & 0 & 30 & 1  & A\^{}1Piu & 0 & 31 & 3 & 88DoNiBeb.125\\
3350.6451 & 0.007 & b\^{}3Sigmag- & 4 & 27 & 2  & a\^{}3Piu & 5 & 27 & 2 & 15ChKaBeTa.1582\\
3351.6007 & 0.003 & B\^{}1Deltag & 1 & 8 & 1 & A\^{}1Piu & 1 & 9 & 1 & 16ChKaBeTa.169\\
3352.8642 & 0.003 & B\^{}1Deltag & 0 & 39 & 1 & A\^{}1Piu & 0 & 39 & 1  & 16ChKaBeTa.170\\
3354.5954 & 0.003 & B\^{}1Deltag & 1 & 15 & 1 & A\^{}1Piu & 1 & 15 & 1 & 16ChKaBeTa.171\\
3356.9124 & 0.007 & b\^{}3Sigmag- & 4 & 14 & 3 & a\^{}3Piu & 5 & 15 & 3 & 15ChKaBeTa.1557\\
3357.1991 & 0.003176643 & B\^{}1Deltag & 1 & 7 & 1  & A\^{}1Piu & 1 & 8 & 1 & 16ChKaBeTa.172\\
\bottomrule
\end{tabular}

\begin{tabular}{ccl}
\\
             Column        &     Notation                  &       \\
\midrule
   1  &    $\tilde{\nu}$              &     Transition frequency (in \cm) \\
   2  &  $\Delta\tilde{\nu}$         &    Estimated uncertainty in transition frequency (in \cm) \\
   3  &   State$^\prime$  &   Electronic state of upper energy level; also includes parity for $\Pi$ states and $\Omega$ for triplet states \\
    4  &  $v^\prime$  &  Vibrational quantum number  of upper  level \\ 
  5  &   $J^\prime$                &        Total angular momentum of upper  level   \\
  6  &   $F^\prime$                &        Spin multiplet component of upper level, labelled in energy order   \\
   7  &   State$^{\prime\prime}$  &   Electronic state of lower energy level; also includes parity for $\Pi$ states and $\Omega$ for triplet states \\
    8  &  $v^{\prime\prime}$  &  Vibrational quantum number  of lower  level \\

   9  &   $J^{\prime\prime}$                &        Total angular momentum of lower  level   \\
    10    &   $F^{\prime\prime}$                &      Spin multiplet component of lower  level, labelled in energy order    \\

   11  &  ID  &  Unique ID for transition, with reference key for source and counting number \\
\bottomrule
\end{tabular}
\end{table*}

The parity-defining quantum numbers used in the original compilation are not necessary given that lambda-doubling transitions have not been observed in $^{12}$C$_2$, and have thus been removed in the present study for simplicity. 

The new transitions file thus has the format shown in \Cref{tab:marvelinput}. 
The transitions file serves as both input to the \Marvel{} procedure and as 
a single consolidated source of assigned transition frequencies and 
uncertainties for \ce{^{12}C2}.  

 Salient details of each source of new and updated experimental data are summarised
 in \Cref{tab:NewOld,tab:NewNew,tab:Updates}, which specify, for each included 
 vibronic band,
 (1) the number of total and validated transitions, 
 (2) the average and maximum uncertainty of the spectral lines after self-consistency, and (3) the $J$ and the wavenumber ranges of the transitions. 
 The same details are provided in the supplementary information for transitions retained from the original \Marvel{}  compilation  \citep{jt637}.

\begin{table}
\footnotesize
    \centering
    \caption{\label{tab:NewOld}Newly included data sources 
    (tags in bold).
    Given are details of the vibronic band, 
    the vibrational states (Vib.) and the total angular momentum
    quantum numbers ($J$) involved, the number of transitions (Trans.)
    validated (V) and original accessed (A), the wavenumber (Wn)
    range of the band, and information about source uncertainties (Unc.), with their
    average (Av) and maximum (Max) values.}
    
\resizebox{\columnwidth}{!}{%
    \begin{tabular}{p{1.8cm}p{0.8cm}p{0.8cm}p{1.cm}p{1.6cm}p{1.5cm}}
    \toprule
Band & Vib.& $J$-range  & Trans. (V/A) & Wn~range (\cm{}) & Unc. (\cm{}) (Av/Max) \\
\midrule
\mc{5}{l}{\textbf{30DiLo}  \citep{30DiLo}} \\ 
 \Cstate{} $-$ \Astate{} & $(0-0)$ & 1--73 & 143/143 & 25\,952-27\,066 & 0.202/0.350  \\ 
\Cstate{} $-$ \Astate{} & $(0-1)$ & 1--86  & 153/158& 24\,370-25\,722 & 0.194/0.200   \\ 
\Cstate{} $-$ \Astate{} & $(1-0)$ & 1--71  & 133/138& 27\,714-28\,617 & 0.205/0.432   \\ 
\Cstate{} $-$ \Astate{} & $(1-1)$ & 6--64  & 107/111& 26\,132-26\,981 & 0.205/0.500   \\ 
\Cstate{} $-$ \Astate{} & $(1-2)$ & 1--72  & 136/136& 24\,574-25\,690 & 0.201/0.349   \\ 
\Cstate{} $-$ \Astate{} & $(2-1)$ & 3--67  & 122/124& 27\,825-28\,575 & 0.202/0.332   \\ 
\Cstate{} $-$ \Astate{} & $(2-3)$ & 7--54  & 83/88& 24\,734-25\,375 & 0.194/0.450   \\ 
\mc{5}{l}{\textbf{37FoHe}  \citep{37FoHe.C2}} \\ 
  \estate{} $-$ \astate{}& $(0-3)$ & 5--43 & 176/182 & 34\,445-35\,025 & 0.310/0.642  \\ 
 \estate{} $-$ \astate{}& $(0-4)$ & 3--47  & 212/230& 32\,798-33\,487 & 0.373/0.684   \\ 
 \estate{} $-$ \astate{}& $(0-5)$ & 22--45  & 87/90& 31\,300-31\,689 & 0.274/0.500   \\ 
 \estate{} $-$ \astate{}& $(0-6)$ & 17--37  & 63/75& 30\,011-30\,301 & 0.247/0.638   \\ 
\mc{5}{l}{\textbf{40HeSu}  \citep{40HeSu} }\\ 
 \Cstate{} $-$ \Astate{} & $(3-1)$ & 1--24 & 45/45 & 29\,419-29\,577 & 0.207/0.291  \\ 
\Cstate{} $-$ \Astate{} & $(4-2)$ & 1--36  & 68/68& 29\,289-29\,504 & 0.200/0.200   \\ 
\Cstate{} $-$ \Astate{} & $(5-3)$ & 1--35  & 56/56& 28\,910-29\,899 & 0.200/0.200   \\ 
\Cstate{} $-$ \Astate{} & $(5-4)$ & 1--32  & 62/62& 27\,439-27\,632 & 0.200/0.200   \\ 
\Cstate{} $-$ \Astate{} & $(6-5)$ & 1--41  & 73/79& 26\,714-27\,100 & 0.185/0.200   \\ 
\mc{5}{l}{\textbf{50Phillips}  \citep{50Phillips} }\\ 
 \Cstate{} $-$ \Astate{} & $(7-6)$ & 1--37 & 64/64 & 26\,954-27\,229 & 0.200/0.200  \\ 
\mc{5}{l}{\textbf{67Messele}  \citep{67Messerle}} \\ 
 \Cstate{} $-$ \Astate{} & $(5-4)$ & 1--53 & 103/103 & 27\,231-27\,632 & 0.029/0.092  \\ 
\mc{5}{l}{\textbf{69HeLaMa}  \citep{69HeLaMa}} \\ 
  \Fstate{} $-$ \Xstate{} & $(0-0)$ & 10--38 & 42/42 & 74\,153-74\,550 & 0.121/0.256  \\ 
 \Fstate{} $-$ \Xstate{} & $(1-0)$ & 2--34  & 38/38& 75\,812-76\,102 & 0.141/0.484   \\ 
 \fstate{} $-$ \astate{}& $(0-0)$ & 3--32  & 111/113& 69\,974-70\,208 & 0.152/0.480   \\ 
 \fstate{} $-$ \astate{}& $(1-0)$ & 2--34  & 113/115& 71\,232-71\,538 & 0.159/0.500   \\ 
 \fstate{} $-$ \astate{}& $(2-0)$ & 2--26  & 95/100& 72\,611-72\,840 & 0.178/0.500   \\ 
 \gstate{} $-$ \astate{}& $(0-0)$ & 3--32  & 202/230& 71\,467-71\,674 & 0.114/0.500   \\ 
 \gstate{} $-$ \astate{}& $(1-0)$ & 3--31  & 199/206& 72\,978-73\,128 & 0.141/0.500   \\ 

\mc{5}{l}{ \textbf{88GoCo}  \citep{88GoCob}} \\ 
  \Onedeltastate{} $-$ \Astate{} & $(0-0)$ & 2--28 & 66/66 & 48\,796-49\,126 & 0.659/1.841  \\ 
 \Onedeltastate{} $-$ \Astate{} & $(0-1)$ & 2--30  & 94/94& 47\,148-47\,547 & 1.032/2.161   \\ 
 \Onedeltastate{} $-$ \Astate{} & $(0-2)$ & 2--33  & 121/121& 45\,604-45\,988 & 0.435/1.475   \\ 
 \Onedeltastate{} $-$ \Astate{} & $(0-3)$ & 2--26  & 67/67& 44\,118-44\,457 & 0.503/1.180   \\ 
 \Onedeltastate{} $-$ \Astate{} & $(0-4)$ & 1--21  & 88/88& 42\,708-42\,949 & 0.702/1.079   \\ 
 \Onedeltastate{} $-$ \Astate{} & $(1-0)$ & 3--34  & 88/88& 49\,850-50\,231 & 0.380/1.144   \\ 
 \Onedeltastate{} $-$ \Astate{} & $(1-1)$ & 2--37  & 125/125& 48\,194-48\,650 & 0.566/1.355   \\ 
 \Onedeltastate{} $-$ \Astate{} & $(1-2)$ & 2--32  & 88/88& 46\,803-47\,093 & 0.418/1.266   \\ 
 \Onedeltastate{} $-$ \Astate{} & $(1-3)$ & 2--28  & 78/79& 45\,210-45\,551 & 0.761/1.342   \\ 
 \Onedeltastate{} $-$ \Astate{} & $(1-4)$ & 3--28  & 48/48& 43\,791-44\,040 & 0.672/1.502   \\ 
 \Onedeltastate{} $-$ \Astate{} & $(2-0)$ & 2--31  & 102/102& 50\,873-51\,287 & 0.399/1.370   \\ 
 \Onedeltastate{} $-$ \Astate{} & $(2-1)$ & 3--28  & 53/53& 49\,538-49\,708 & 1.087/1.879   \\ 
 \Onedeltastate{} $-$ \Astate{} & $(2-4)$ & 2--19  & 56/56& 44\,926-45\,111 & 0.873/1.643   \\ 

\mc{5}{l}{\textbf{89GoCo}  \citep{89GoCo}}\\ 
\Onedeltastate{} $-$ \Bstate{}  & $(0-1)$ & 2--29 & 60/60 & 43\,973-44\,140 & 0.658/1.223  \\
 \bottomrule
\end{tabular}
}
\end{table}

\begin{table}
    \centering
    \footnotesize
    \caption{\label{tab:NewNew}New experimental data sources for $^{12}$C$_2$ 
    which appeared since the original 
    \Marvel{} study  \citep{jt637} was published. Details as in \Cref{tab:NewOld}.}
    
\resizebox{\columnwidth}{!}{%

    \begin{tabular}{p{1.9cm}p{1cm}p{0.8cm}p{1.1cm}p{1.6cm}p{1.5cm}}
    \toprule
Band & Vib. & $J$-range  & Trans. (V/A) & Wn~range (\cm{}) & Unc. (\cm{}) (Av/Max) \\
\midrule
\mc{5}{l}{\textbf{17BoViBeKn}  \citep{17BoViBeKn}} \\ 
  $1 {}^5\Pi_g-$ \astate{}& $(3-6)$ & 4-4 & 1/3 & 22\,733-22\,733 & 0.018/0.055  \\ 
 $1 {}^5\Pi_u - 1 {}^5\Pi_g$ & $(1-0)$ & 1-6  & 11/11& 22\,601-22\,632 & 0.055/0.055   \\ 
\dstate{} $-$ \astate{}& $(8-6)$ & 1-10  & 39/41& 22\,722-22\,774 & 0.092/0.302   \\ 

\mc{5}{l}{\textbf{17WeKrNaBa}  \citep{17WeKrNaBa}} \\ 
  \estate{} $-$ \astate{}& $(0-2)$ & 1-17 & 112/112 & 36\,472-36\,641 & 0.047/0.208  \\ 
 \estate{} $-$ \astate{}& $(1-3)$ & 1-13  & 118/118& 35\,969-36\,105 & 0.067/0.258   \\ 
 \estate{} $-$ \astate{}& $(10-0)$ & 0-12  & 67/67& 48\,331-48\,461 & 0.060/0.309   \\ 
 \estate{} $-$ \astate{}& $(2-2)$ & 1-13  & 80/80& 38\,536-38\,652 & 0.078/0.194   \\ 
 \estate{} $-$ \astate{}& $(2-3)$ & 1-15  & 115/115& 36\,932-37\,082 & 0.058/0.222   \\ 
 \estate{} $-$ \astate{}& $(3-4)$ & 1-13  & 124/124& 36\,316-36\,466 & 0.054/0.264   \\ 
 \estate{} $-$ \astate{}& $(4-0)$ & 1-16  & 160/160& 43\,486-43\,689 & 0.060/0.213   \\ 
 \estate{} $-$ \astate{}& $(5-0)$ & 0-15  & 131/131& 44\,346-44\,550 & 0.048/0.257   \\ 
 \estate{} $-$ \astate{}& $(5-3)$ & 0-8  & 51/51& 39\,697-39\,767 & 0.070/0.188   \\ 
 \estate{} $-$ \astate{}& $(6-0)$ & 0-16  & 116/116& 45\,210-45\,381 & 0.066/0.239   \\ 
 \estate{} $-$ \astate{}& $(6-3)$ & 1-10  & 43/43& 40\,526-40\,598 & 0.091/0.249   \\ 
 \estate{} $-$ \astate{}& $(7-0)$ & 0-12  & 94/94& 46\,030-46\,186 & 0.053/0.184   \\ 
 \estate{} $-$ \astate{}& $(8-0)$ & 1-15  & 121/121& 46\,754-46\,967 & 0.056/0.189   \\ 
 \estate{} $-$ \astate{}& $(9-0)$ & 0-14  & 98/98& 47\,570-47\,724 & 0.056/0.160   \\ 

\mc{5}{l}{\textbf{17KrWeBa}  \citep{17KrWeBaNa}}\\ 
  \threestate{} $-$ \astate{}& $(3-2)$ & 2-7 & 29/29 & 46\,842-46\,894 & 0.108/0.160  \\ 
 \threestate{} $-$ \astate{}& $(3-5)$ & 1-8  & 41/41 & 42\,183-42\,252 & 0.116/0.313   \\ 
 \threestate{} $-$ \astate{}& $(8-3)$ & 2-7  & 9/9 & 49\,181-49\,205 & 0.180/0.373   \\ 
 \threestate{} $-$ \astate{}& $(8-4)$ & 1-6  & 19/19 & 47\,596-47\,655 & 0.170/0.286   \\ 
 \threestate{} $-$ \astate{}& $(8-5)$ & 2-6  & 11/11 & 46\,084-46\,133 & 0.110/0.136   \\ 
 \threestate{} $-$ \astate{}& $(8-6)$ & 1-9  & 24/24 & 44\,583-44\,633 & 0.238/0.456   \\ 
 \fourstate{} $-$ \astate{}& $(0-2)$ & 2-7  & 17/17 & 48\,272-48\,327 & 0.230/0.490   \\ 
 \fourstate{} $-$ \astate{}& $(0-3)$ & 2-8  & 24/24 & 46\,683-46\,756 & 0.184/0.351   \\ 
 \fourstate{} $-$ \astate{}& $(0-4)$ & 2-7  & 13/14 & 45\,165-45\,208 & 0.238/0.345   \\ 
 \fourstate{} $-$ \astate{}& $(0-5)$ & 1-9  & 20/20 & 43\,617-43\,686 & 0.220/0.447   \\ 
 \fourstate{} $-$ \astate{}& $(0-6)$ & 1-8  & 13/13 & 42\,137-42\,184 & 0.147/0.340   \\ 
 \fourstate{} $-$ \astate{}& $(1-2)$ & 3-8  & 16/21 & 49\,544-49\,593 & 0.213/0.349   \\ 
 \fourstate{} $-$ \astate{}& $(1-3)$ & 2-7  & 18/18 & 47\,973-48\,023 & 0.197/0.421   \\ 
 \fourstate{} $-$ \astate{}& $(1-4)$ & 2-6  & 12/13 & 46\,425-46\,473 & 0.317/0.500   \\ 
 \fourstate{} $-$ \astate{}& $(1-5)$ & 2-7  & 20/26 & 44\,900-44\,951 & 0.121/0.297   \\ 
 \fourstate{} $-$ \astate{}& $(2-3)$ & 1-7  & 25/25 & 49\,257-49\,328 & 0.147/0.302   \\ 
 \fourstate{} $-$ \astate{}& $(2-5)$ & 2-7  & 21/21 & 46\,220-46\,270 & 0.127/0.288   \\ 

\mc{5}{l}{\textbf{18KrWeFrNa}  \citep{18KrWeFrNa}} \\ 
 \Dstate{} $-$ \Xstate{} & $(4-2)$ & 1-23  & 17/17& 46\,537-46\,667 & 0.020/0.020   \\ 
 \Dstate{} $-$ \Xstate{} & $(5-3)$ & 1-19  & 15/15& 46\,519-46\,611 & 0.022/0.032   \\ 
 \Dstate{} $-$ \Xstate{} & $(6-4)$ & 1-11  & 12/12& 46\,431-46\,511 & 0.035/0.083   \\ 
 \Dstate{} $-$ \Xstate{} & $(7-5)$ & 1-11  & 12/12& 46\,355-46\,433 & 0.039/0.074   \\ 
 \Dstate{} $-$ \Xstate{} & $(8-6)$ & 1-11  & 9/13& 46\,291-46\,361 & 0.024/0.030   \\ 
 \Dstate{} $-$ \Xstate{} & $(9-7)$ & 1-13  & 12/12& 46\,228-46\,302 & 0.020/0.020   \\ 
  \Dstate{} $-$ \Xstate{} & $(10-8)$ & 1-11 & 12/12 & 46\,167-46\,243 & 0.021/0.025  \\ 
 \Dstate{} $-$ \Xstate{} & $(11-9)$ & 1-11  & 12/12& 46\,130-46\,205 & 0.021/0.028   \\ 

\mc{5}{l}{\textbf{19Nakajima}  \citep{19Nakajima}} \\ 
 \Astate{} $-$ \Xstate{} & $(10-4)$ & 1-30 & 43/43 & 15\,588-15\,882 & 0.005/0.022  \\ 
\Astate{} $-$ \Xstate{} & $(10-5)$ & 1-25  & 35/35& 13\,991-14\,175 & 0.005/0.022   \\ 
\Astate{} $-$ \Xstate{} & $(11-4)$ & 1-21  & 26/26& 17\,103-17\,220 & 0.005/0.017   \\ 
\Astate{} $-$ \Xstate{} & $(11-5)$ & 1-23  & 29/29& 15\,379-15\,513 & 0.005/0.022   \\ 
\Astate{} $-$ \Xstate{} & $(6-0)$ & 1-28  & 40/40& 17\,152-17\,417 & 0.017/0.132   \\ 
\Astate{} $-$ \Xstate{} & $(6-1)$ & 1-26  & 36/36& 15\,378-15\,590 & 0.012/0.130   \\ 
\Astate{} $-$ \Xstate{} & $(9-4)$ & 1-31  & 43/43& 14\,236-14\,520 & 0.004/0.012   \\ 
\bottomrule
\end{tabular}
}
\end{table}

\subsection{New data sources, pre-2016}
The 2016 original \MARVEL{} compilation of \ce{^{12}C2} spectroscopic data
did not incorporate sources that either 
(a) did not include an estimate of the transition frequency uncertainty, or 
(b) involved very high-lying electronic states, \emph{e.g.}, 
\Cstate{}, \Fstate{}, \fstate{}, and \gstate{}. 


Many of these new data sources include transitions in the ultra-violet and have much higher uncertainties than other data sources. 
In most data sources, we limited the maximum uncertainty to 0.5 \cm{}.
Next, we provide information about these sources one by one.

\begin{description}
\item[\textbf{30DiLo}] \citep{30DiLo}
An early source of Deslandres--d'Azambuja (\Cstate{} $-$ \Astate) transitions 
for low-energy vibrational states, which have surprisingly not been remeasured 
for \ce{^{12}C2} despite the high uncertainties of these data.  
Uncertainties were estimated using combination differences based on the use of
lower-state energy levels determined by \MARVEL{},
the estimated source uncertainties are 0.2 \cm{}. 
\item[\textbf{37FoHe}] \citep{37FoHe.C2}
The first measurements of the Fox--Herzberg (\estate{} $-$ \astate{}) band, 
all with upper vibrational state of $v=0$. 
Uncertainties were estimated using combination differences based on the use of 
lower-state energies determined by \MARVEL{}, the estimated 
source uncertainties are 0.2 \cm{}.
Two of these bands, (0--3) and (0--6), have not been remeasured, 
though all involved energy levels are now well understood by other band measurements. 
\item[\textbf{40HeSu}] \citep{40HeSu}
Another source of Deslandres--d'Azambuja (\Cstate{} $-$ \Astate{}) transitions
with a number of vibrational bands.  
Uncertainties were estimated using combination differences based on the use of 
lower-state energies determined by \MARVEL{}, the estimated source
uncertainties are 0.2 \cm{}. 
\item[\textbf{50Phillips}] \citep{50Phillips}
Even another source of Deslandres--d'Azambuja 
(\Cstate{} $-$ \Astate{}) transitions, with significantly excited 
vibrational states involved.
\item[\textbf{67Messerle}] \citep{67Messerle}
Some data from Messerle's 1967 thesis was located  \citep{Klaas}; 
these data are relevant for the Deslandres--d'Azambuja (\Cstate{} $-$ \Astate{}) band. 
Note that no data from the putative Messerle--Krauss bands 
(\Cprimestate{} $-$ \Astate{}) were found. 
\item[\textbf{69HeLaMa}] \citep{69HeLaMa}
These data are the sole source for many very high-lying bands, 
located above 70\,000 \cm{}. 
Uncertainties were estimated using combination differences based on the use of 
lower-state energies determined by \MARVEL{}, the estimated source uncertainties
are 0.1 \cm{}, with blended lines given a starting uncertainty of 0.2 \cm{}.  
Lines with recommended \Marvel{} uncertainties higher than 0.5 \cm{} 
were not validated; this process only removed a small number of lines 
and all bands mostly consisted of validated lines.
\item[\textbf{88GoCo}] \citep{88GoCob}
To ensure reliable inclusion of the 88GoCo  \citep{88GoCob}, and in fact
89GoCo  \citep{89GoCo}, data 
(the only source of data for the \Onedeltastate{} state),
uncertainties up to 2 \cm{} were required to be permitted in these two sources.
These high uncertainty values can be attributed to the quality of the
early ultraviolet studies, which have a stated absolute energy error of 2 \cm{}, 
though the relative uncertainties were stated as $\pm$ 0.2 \cm{}. 
\item[\textbf{89GoCo}] \citep{89GoCo}
Following the considerations of 88GoCo \citep{88GoCob},
an earlier paper by the same authors, 
uncertainties for the 89GoCo data were initially set as 0.2 \cm{}, 
but allowed to increase as required; 
all transitions were validated with uncertainties less than 1.2 \cm{}.
\end{description}

\subsection{New data sources, post-2016}

\begin{description}
\item[\textbf{17BoViBeKn}] \citep{17BoViBeKn}
This paper provides new data on the quintet bands and three assignments of the quintet-triplet spin-forbidden bands. 
However, two of these three new spin-forbidden bands were inconsistent 
with the rest of the \Marvel{} compilation without uncertainties of around 
2.4 \cm{} and were thus excluded; 
in contrast, 67 spin-forbidden transitions between the same two electronic states
in 11BoSyKnGe \citep{11BoSyKnGe.C2} all validated with smaller uncertainties. 

\item[\textbf{17WeKrNaBa}] \citep{17WeKrNaBa}
Rotationally cool experimental conditions enabled detailed study of 
low-$J$ ultraviolet rovibronic transitions in the Fox--Herzberg
(\estate{} $-$ \astate{})  band.
No explicit uncertainty was provided in this paper; 
anestimated source uncertainty of 0.035 \cm{} was used, though this seems to be 
a slight underestimation based on the uncertainties \Marvel{} found. 

\item[\textbf{17KrWeBaNa}] \citep{17KrWeBaNa}
This is the first study of the \threestate{} state,
observed \emph{via} the \threestate{} $-$ \astate{} transitions, 
with the paper also substantially expanding on previously known data 
on the \fourstate{} state.
The estimated source uncertainties are 0.035 \cm{}. 
Detailed \Marvel{}-based analyses of the data revealed that the original assignments
were not self-consistent within this paper.
Following these analyses, one of the original authors identified \citep{Klaas} 
a calibration error in the \fourstate{} $-$ \astate{} ($0-5$) transition 
frequencies, which can be corrected by decreasing all frequencies 
in this band by 0.9 \cm{}.
Though it could not be confirmed, a calibration error was also suspected in the \fourstate{} $-$ \astate{} (1$-$3) transition frequencies; 
the \Marvel{} analysis showed the data set became self-consistent without 
unreasonably large uncertainties if these transition frequencies  
were decreased by 1.0 \cm{}. 
\item[\textbf{18KrWeFrNa}] \citep{18KrWeFrNa}
Rotationally cool experimental conditions enabled the detailed study of low-$J$
rovibronic transitions in the  
Mulliken (\Dstate{} -- \Xstate{}) $\Delta v = \pm 2$ sequence. 
This paper does not present explicit uncertainties for the spectral lines.
Given the ultraviolet frequency of these transitions, an estimated source
uncertainty of 0.02 \cm{} was assigned, with reasonable results. 
\item[\textbf{19Nakajima}] \citep{19Nakajima}
The paper provides an estimate for the line uncertainties as 0.01 \cm{}, 
which we adopted as the estimated source uncertainty. 
\end{description}



\begin{table}
\footnotesize
    \centering
    \caption{\label{tab:Updates}Updates to previously included data sources.  Details as in \Cref{tab:NewOld}.}
    \resizebox{\columnwidth}{!}{%

    \begin{tabular}{p{1.9cm}p{1cm}p{0.8cm}p{1.1cm}p{1.6cm}p{1.5cm}}
    \toprule
Band & Vib. & $J$-range  & Trans. (V/A) & Wn range (\cm{}) & Unc. (\cm{}) (Av/Max) \\
\midrule
\mc{5}{l}{\emph{48Phillips} (minor corrections)   \citep{48Phillips.C2a} }\\
\dstate{} $-$ \astate{}& $(0-2)$ & 1-53 & 247/247& 16152-16751 & 0.056/0.275  \\ 
\dstate{} $-$ \astate{}& $(1-3)$ & 2-56  & 228/230& 16332-16690 & 0.066/0.288   \\ 
\dstate{} $-$ \astate{}& $(10-9)$ & 0-43  & 194/196& 20697-20957 & 0.065/0.297   \\ 
\dstate{} $-$ \astate{}& $(2-4)$ & 11-42  & 153/153& 16498-16864 & 0.079/0.254   \\ 
\dstate{} $-$ \astate{}& $(3-5)$ & 1-30  & 110/110& 16662-16868 & 0.073/0.338   \\ 
\dstate{} $-$ \astate{}& $(8-6)$ & 5-32  & 114/117& 22648-22812 & 0.074/0.414   \\ 
\dstate{} $-$ \astate{}& $(9-8)$ & 15-43  & 124/128& 20962-21198 & 0.080/0.268   \\ 

\mc{5}{l}{\emph{49Phillips} (re-processed)   \citep{49Phillips.C2} }\\ 
  \estate{} $-$ \astate{}& $(0-2)$ & 6-39 & 166/172& 35963-36596 & 0.144/0.632  \\ 
 \estate{} $-$ \astate{}& $(1-2)$ & 4-55  & 234/266& 36506-37635 & 0.099/0.709   \\ 
 \estate{} $-$ \astate{}& $(1-3)$ & 3-52  & 226/259& 35091-36064 & 0.129/0.728   \\ 
 \estate{} $-$ \astate{}& $(2-1)$ & 5-43  & 198/203& 39453-40202 & 0.100/0.746   \\ 
 \estate{} $-$ \astate{}& $(2-2)$ & 4-49  & 234/235& 37643-38608 & 0.081/0.534   \\ 
 \estate{} $-$ \astate{}& $(2-3)$ & 8-35  & 129/129& 36618-37032 & 0.120/0.440   \\ 
 \estate{} $-$ \astate{}& $(3-1)$ & 4-41  & 185/188& 40429-41127 & 0.101/0.652   \\ 
 \estate{} $-$ \astate{}& $(3-2)$ & 15-48  & 148/149& 38595-39370 & 0.103/0.654   \\ 
 \estate{} $-$ \astate{}& $(4-1)$ & 5-44  & 196/197& 41174-42019 & 0.096/0.611   \\ 

\mc{5}{l}{\emph{86HaWi} (extended)  \citep{86HaWi} }\\ 
  \estate{} $-$ \astate{}& $(0-4)$ & 2-39 & 192/197 & 33035-33502 & 0.174/0.500  \\ 
 \estate{} $-$ \astate{}& $(0-5)$ & 2-38  & 219/221& 31386-31977 & 0.098/0.413   \\ 
 \estate{} $-$ \astate{}& $(1-8)$ & 2-31  & 171/179& 28195-28580 & 0.104/0.446   \\ 



\mc{5}{l}{\emph{07TaHiAm} (re-processed)  \citep{07TaHiAm.C2}} \\ 
\dstate{} $-$ \astate{}& $(0-0)$ & 1-87 & 411/417& 19355-20602 & 0.005/0.047  \\ 
\dstate{} $-$ \astate{}& $(0-1)$ & 0-38  & 191/191& 17740-18085 & 0.007/0.030   \\ 
\dstate{} $-$ \astate{}& $(0-2)$ & 1-38  & 171/171& 16147-16514 & 0.006/0.018   \\ 
\dstate{} $-$ \astate{}& $(1-0)$ & 0-53  & 214/215& 21104-21587 & 0.008/0.041   \\ 
\dstate{} $-$ \astate{}& $(1-1)$ & 0-52  & 174/174& 19490-19999 & 0.008/0.041   \\ 
\dstate{} $-$ \astate{}& $(1-2)$ & 0-29  & 138/138& 17899-18117 & 0.008/0.040   \\ 
\dstate{} $-$ \astate{}& $(10-9)$ & 1-36  & 160/160& 20749-20958 & 0.005/0.042   \\ 
\dstate{} $-$ \astate{}& $(2-0)$ & 2-34  & 118/118& 22814-23044 & 0.014/0.075   \\ 
\dstate{} $-$ \astate{}& $(2-1)$ & 0-40  & 84/86& 21202-21514 & 0.007/0.022   \\ 
\dstate{} $-$ \astate{}& $(2-2)$ & 3-36  & 51/51& 19612-19910 & 0.007/0.053   \\ 
\dstate{} $-$ \astate{}& $(2-3)$ & 0-37  & 63/63& 18043-18349 & 0.009/0.040   \\ 
\dstate{} $-$ \astate{}& $(3-1)$ & 1-35  & 170/170& 22869-23109 & 0.025/0.117   \\ 
\dstate{} $-$ \astate{}& $(3-2)$ & 2-39  & 159/161& 21282-21573 & 0.012/0.054   \\ 
\dstate{} $-$ \astate{}& $(3-3)$ & 1-30  & 139/139& 19715-19933 & 0.015/0.096   \\ 
\dstate{} $-$ \astate{}& $(3-4)$ & 1-29  & 40/41& 18172-18359 & 0.009/0.061   \\ 
\dstate{} $-$ \astate{}& $(3-5)$ & 0-31  & 137/137& 16648-16846 & 0.007/0.039   \\ 
\dstate{} $-$ \astate{}& $(3-6)$ & 0-30  & 129/130& 15149-15390 & 0.008/0.055   \\ 
\dstate{} $-$ \astate{}& $(4-3)$ & 2-43  & 119/119& 21339-21647 & 0.008/0.096   \\ 
\dstate{} $-$ \astate{}& $(4-5)$ & 3-25  & 65/65& 18276-18441 & 0.011/0.081   \\ 
\dstate{} $-$ \astate{}& $(4-6)$ & 3-25  & 66/66& 16777-16968 & 0.011/0.093   \\ 
\dstate{} $-$ \astate{}& $(5-4)$ & 0-37  & 191/192& 21368-21615 & 0.015/0.102   \\ 
\dstate{} $-$ \astate{}& $(5-5)$ & 20-20  & 1/1& 19984-19984 & 0.005/0.005   \\ 
\dstate{} $-$ \astate{}& $(5-6)$ & 0-32  & 88/88& 18353-18584 & 0.007/0.027   \\ 
\dstate{} $-$ \astate{}& $(5-7)$ & 1-20  & 42/42& 16879-17004 & 0.006/0.014   \\ 
\dstate{} $-$ \astate{}& $(6-5)$ & 0-37  & 136/138& 21360-21594 & 0.007/0.049   \\ 
\dstate{} $-$ \astate{}& $(6-8)$ & 0-26  & 52/52& 16946-17129 & 0.007/0.053   \\ 
\dstate{} $-$ \astate{}& $(7-6)$ & 2-37  & 135/136& 21297-21523 & 0.009/0.178   \\ 
\dstate{} $-$ \astate{}& $(7-9)$ & 1-25  & 93/93& 16970-17121 & 0.007/0.031   \\ 
\dstate{} $-$ \astate{}& $(8-6)$ & 3-26  & 91/91& 22666-22824 & 0.009/0.063   \\ 
\dstate{} $-$ \astate{}& $(8-7)$ & 0-14  & 35/36& 21228-21311 & 0.012/0.085   \\ 
\dstate{} $-$ \astate{}& $(9-8)$ & 0-37  & 172/172& 20990-21208 & 0.006/0.050   \\ 

\mc{5}{l}{\emph{16ChKaBeTa} (new lines)  \citep{16ChKaBeTa.C2}} \\ 
  \Bprimestate{} $-$ \Astate{} & $(4-3)$ & 0-28 & 35/35 & 7798-7949 & 0.004/0.013  \\ 
 \Bprimestate{} $-$ \Astate{} & $(4-5)$ & 2-30  & 39/39& 4790-4958 & 0.004/0.024   \\ 
 \Bprimestate{} $-$ \Astate{} & $(4-6)$ & 2-26  & 30/30& 3400-3501 & 0.006/0.013   \\ 
\bottomrule
\end{tabular}
}
\end{table}

\subsection{Corrections to the 2016 compilation}
During the process of updating the compilation of $^{12}$C$_2$ rovibronic data,
a number of issues with the original data were identified and corrected. 
A source-by-source specification of the corrections follows. 
\renewcommand{\descriptionlabel}[1]{\hspace{\labelsep}\textbf{#1}:}

\begin{description}
\item[\emph{48Phillips}] \citep{48Phillips.C2b}
A small number of digitisation errors were identified and corrected following 
a thorough re-examination of the band structure. 
\item[\emph{49Phillips}] \citep{49Phillips.C2}
Significant errors with quantum numbers, including incorrect band assignments, 
were identified and corrected, and the repetition of one band's data 
identified and removed. 
\item[\emph{86HaWi}] \citep{86HaWi}
The 2016 compilation included only 100 of the spectral lines reported in this paper; 
the other 497 lines have been added to this update. 
\item[\emph{07TaHiAm}] \citep{07TaHiAm.C2}
The original data compilation included 3813 transitions, 
but some of these transitions were  calculated rather than measured and 
many measured transitions were excluded.  
The source was reprocessed into \Marvel{} format, 
giving a total of 4813 transitions, with calculated transitions (labelled ``z'') 
excluded, blended lines given a starting uncertainty of 0.01 \cm{} and 
well resolved isolated lines given an estimated source uncertainty of 0.005 \cm{}. 
Note that the data from the 02TaAm \citep{02TaAm.C2} source 
that was reproduced in 07TaHiAm was not included.  
Given the high resolution of the data, transitions with  uncertainties 
greater than 0.2 \cm{} were deemed misassignments and not validated 
(though they are still in the included \Marvel{} file with a "-" in front 
of the transition frequency, as usual); 
in total 88 transitions were not validated, mostly at the beginning 
or end of a given band. 
When processing these data to determine optimal uncertainties, 
the 48Phillips \citep{48Phillips.C2b} and 49Phillips \citep{49Phillips.C2} data \edited{were} removed before being re-added; this ensured the more recent data had 
a larger number of valid transitions with lower uncertainties than the older data.
\item[\emph{16ChKaBeTa}] \citep{16ChKaBeTa.C2}
Due to late inclusion of this data source, 
the lines from the \Bprimestate{} -- \Astate{} band were missed during
the original compilation; these 104 missing lines have been added in this update.  
\end{description}

Additionally, for 06PeSi \citep{06PeSi}, we want to note that we retained the 2016 \Marvel{} values, but that these were reassigned from the original paper's assignments because the original contained unphysical assignments for the quantum numbers of homonuclear diatomics.

\subsection{Discussion} 

\begin{figure*}
    \centering
    \includegraphics[width=1\textwidth]{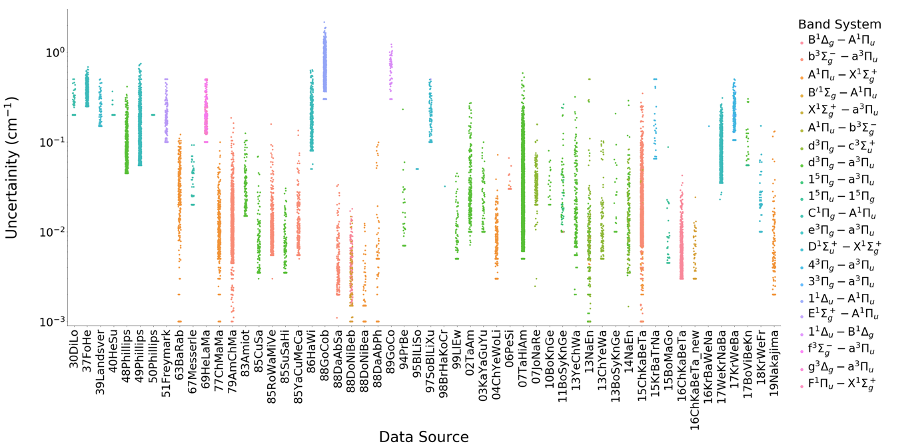}
    \caption{The range of \edited{transition} frequency uncertainties for every source involved in the full \Marvel{} input file. See \Cref{tab:bands} for the citations to these sources.}
    \label{fig:source-unc}
\end{figure*}

The spread of uncertainties for every data source used for the updated 
\Marvel{} input is shown in \Cref{fig:source-unc}. 
The horizontal axis goes through each data source in chronological order.
The colours of the data are assigned to the electronic bands of the transition, ordered by transitions frequency. 
This plot clearly shows that for each source there are a significant number of
transitions with uncertainties above the minimum uncertainty for that data set;
these uncertainty increases were required to ensure self-consistency with the 
data coming from other sources. 
The figure makes it also clear that pre-1960 sources have much larger uncertainties
than the later data, though there is no clear trend in improved accuracy since 1960. 
Most data sources since 1960 have minimum uncertainties between 10$^{-3}$ 
to 10$^{-2}$ \cm{}. 
The post-1960 outliers with higher uncertainties, on the order of 0.1 \cm{}, are usually very-high-frequency transitions in the ultraviolet region. These higher uncertainties are expected as this region is more spectrally congested and UV instruments are more expensive with lower spectral resolution than visible instruments due to decreased market demand. 

Colours in \Cref{fig:PES}, \emph{vide supra},
demonstrate that the new data sources added to the 
$^{12}$C$_2$ \Marvel{} compilation in this paper increase the number of bands 
considered from 11 to 21, with updates in data available for 7 of the 
originally included bands. 
The specific experimental sources in our \Marvel{} compilation for each electronic transition band system are detailed in \Cref{tab:bands}. It is clear that some band systems, especially the \bstate{} $-$ \astate{}, \dstate{} $-$ \astate{}, \estate $-$ \astate, \Astate $-$ \Xstate{} ones, have been extensively explored with thousands of measured transitions in up to 16 different publications. These heavily explored band systems are strong absorption bands in astrophysical environments. In contrast, some other systems have data only from a single paper. 
The triplet manifold is generally better explored than the singlet manifold, 
despite the fact that the electronic ground state of C$_2$ is a singlet. 
The number of quintet and spin-forbidden intersystem lines remains quite small,
but the latter in particular are very important for setting the relative energies 
of the singlet, triplet, and quintet manifolds accurately. 

\Cref{tab:bands} also provides details about the number of 
unique versus total transitions measured. 
The high number of unique versus total transitions in these data clearly 
demonstrates that most re-examinations of a particular band system produce 
data for different vibronic bands rather than re-measuring existing data 
at a higher accuracy. 
This result emphasises the need for a centrally-collated source of all available experimental data in one consistent format, as provided by this paper.

\begin{table*}
    \caption{ \label{tab:bands} All experimental data sources of 
    rotationally-resolved assigned transitions (Trans.) 
    for all investigated band system in $^{12}$C$_2$, where Tot refers to the total number of transitions and Uniq refers to the number of unique transitions. 
    Bold band names indicate bands newly included in this update, 
    while italicized names indicate pre-existing bands with new data. 
    Bold sources are newly included in this 2020 \Marvel{} update, italicised sources are updated from the previous 2016 \Marvel{} compilation. 
    }
    \centering

\resizebox{2\columnwidth}{!}{%
    \begin{tabular}{r@{\hskip 0.1in}p{2cm}cp{10cm}}
\toprule
Multiplicity~~~~~~~~Band name & Band System & Trans. (Tot/Uniq) & Sources \\
\midrule
\mc{2}{l}{\textbf{\underline{Singlet}}} \\
\emph{Phillips} & \Astate{} $-$ \Xstate{}  & 2729/2248 &   63BaRab \citep{63BaRab.C2}, 77ChMaMa \citep{77ChMaMa.C2}, 88DaAbPh \citep{88DaPhRaAb}, 88DoNiBea \citep{88DoNiBea.C2}, 04ChYeWoLi \citep{04ChYeWoLi}, 06PeSi \citep{06PeSi}, 13NaEn \citep{13NaEn.C2}, 15ChKaBeTa \citep{15ChKaBeTa.C2}, \textbf{19Nakajima} \citep{19Nakajima}\\

Bernath B &  \Bstate{} $ -$ \Astate{}& 1508/1508 &   88DoNiBeb \citep{88DoNiBeb.C2}, 16ChKaBeTa \citep{16ChKaBeTa.C2} \\

\emph{Bernath B'} &\Bprimestate{} $-$ \Astate{}& 341/341 &   88DoNiBeb \citep{88DoNiBeb.C2}, \emph{16ChKaBeTa} \citep{16ChKaBeTa.C2} \\

\textbf{Deslandres--d'Azambuja} &\Cstate{} $-$ \Astate{} & 1375/1375 &    \textbf{30DiLo} \citep{30DiLo}, \textbf{40HeSu} \citep{40HeSu}, \textbf{50Phillips} \citep{50Phillips}, \textbf{67Messerle} \citep{67Messerle} \\

\emph{Mulliken} &  \Dstate{} $ -$ \Xstate{} &  464/299 & 39Landsver \citep{39Landsver.C2}, 95BlLiSo \citep{95BlLiSo.C2}, 97SoBlLiXu \citep{97KaHuEw}, \textbf{18KrWeFrNa} \citep{18KrWeFrNa} \\

Freymark & \Estate{} $ -$ \Astate{}&  442/442 & 51Freymark \citep{51Freymark.C2}, 97SoBlLiXu \citep{97SoBlLiXu} \\

\textbf{Herzberg F} & \Fstate{}$-$ \astate{}  & 80/80 &  \textbf{69HeLaMa} \citep{69HeLaMa} \\

\textbf{Goodwin--Cool} &$1~{}^1\Delta_u -$ \Astate{}&   1075/1075 & \textbf{88GoCo} \citep{88GoCob} \\

\textbf{Goodwin--Cool} &$1~{}^1\Delta_u -$ \Bstate{} &   60/60 & \textbf{89GoCo} \citep{89GoCo} \\
\vspace{-0.5em} \\

\mc{2}{l}{\textbf{\underline{Triplet}}} \\
Ballik--Ramsay & \bstate{} $ -$ \astate{}  & 7813/5739 &   79AmChMa \citep{79AmChMa.C2}, 85RoWaMiVe \citep{85RoWaMiVe.C2}, 85YaCuMeCa \citep{85YaCuMeCa}, 88DaAbSa \citep{88DaPhRaAb}, 06PeSi \citep{06PeSi}, 11BoSyKnGe \citep{11BoSyKnGe.C2}, 15ChKaBeTa \citep{15ChKaBeTa.C2}\\

\emph{Swan} & \dstate{} $-$ \astate{}  & 8518/6585 &  48Phillips \citep{48Phillips.C2a}, 83Amiot \citep{83Amiot}, 85CuSa \citep{85CuSa}, 85SuSaHi \citep{85SuSaHi.C2}, 94PrBe \citep{94PrBe.C2}, 99LlEw \citep{99LlEw}, 02TaAm \citep{02TaAm.C2}, 03KaYaGuYu \citep{03KaYaGuYu}, 07TaHiAm \citep{07TaHiAm.C2}, 10BoKnGe \citep{10BoKnGe.C2}, 11BoSyKnGe \citep{11BoSyKnGe.C2}, 13BoSyKnGe \citep{13BoSyKnGe}, 13NaEn \citep{13NaEn.C2}, 13YeChWa \citep{13YeChWa.C2}, 14NaEn \citep{14NaEn.C2}, \textbf{17BoViBeKn} \citep{17BoViBeKn} \\

Duck & \dstate{} $-$ \cstate{} & 1174/985 & 07JoNaRe \citep{07JoNaRe.C2}, 13ChYeWa \citep{13ChYeWa.C2}, 13NaEn \citep{13NaEn.C2}, 14NaEn \citep{14NaEn.C2} \\

\emph{Fox-Herzberg} & \estate{} $-$ \astate{}  &   4412/4085 & \textbf{37FoHe} \citep{37FoHe.C2}, \emph{49Phillips \citep{49Phillips.C2}}, \emph{86HaWi} \citep{86HaWi}, 98BrHaKoCr \citep{98BrHaKoCr}, \textbf{17WeKrNaBa} \citep{17WeKrNaBa} \\

\textbf{Herzberg f} & \fstate{}$-$ \astate{}  & 328/328 &  \textbf{69HeLaMa} \citep{69HeLaMa} \\

\textbf{Herzberg g} & \gstate{}$-$ \astate{}  & 436/436 &  \textbf{69HeLaMa} \citep{69HeLaMa}\\

\textbf{Krechkivska-Schmidt} & \threestate{} $-$ \astate{} &  133/133 &  \textbf{17KrWeBaNa}  \citep{17KrWeBaNa} \\

\emph{Krechkivska--Schmidt} & \fourstate{} $-$ \astate{} &  280/259 & 15KrBaTrNa \citep{15KrBaTrNa} 16KrBaWeNa \citep{16KrBaWeNa} \textbf{17KrWeBaNa} \citep{17KrWeBaNa} \\
\vspace{-0.5em} \\

\mc{2}{l}{\textbf{\underline{Quintet}}} \\
\emph{Radi--Bornhauser} & $1 {}^5\Pi_u - 1 {}^5\Pi_g $&  68/63 & 15BoMaGo \citep{15BoMaGo}, \textbf{17BoViBeKn} \citep{17BoViBeKn} \\
\vspace{-0.5em} \\

\mc{2}{l}{\textbf{\underline{Intercombination}}} \\
& \Xstate{} $-$ \astate{} & 16/16 &  15ChKaBeTa \citep{15ChKaBeTa.C2} \\
& $1~{}^5\Pi_g -$ \astate{} &  70/70 & 11BoSyKnGe \citep{11BoSyKnGe.C2}, \textbf{17BoViBeKn} \citep{17BoViBeKn} \\
& \Astate{} $-$ \bstate{} & 1/1 & 15ChKaBeTa \citep{15ChKaBeTa.C2} \\
\bottomrule

    \end{tabular}}
\end{table*}

With these new included data, every observed band has \Marvel-\edited{compiled} rotationally-resolved data with two exceptions: 
the Kable--Schmidt \citep{09NaJoPaRe.C2} \estate{} $-$ \cstate{} band 
around 40\,000 \cm{} and the Messerle--Krauss \citep{67MeKrxx.C2} \Cprimestate{} $-$ \Astate{} band around 30\,000 \cm{}.
Initial errors in the e state constants prohibited a good fit to the Kable--Schmidt band in 2009; new constants \citep{17WeKrNaBa} obtained from better data in the Fox--Herzberg band allowed a much better fit of the \estate{} $-$ \cstate{} (4$-$3) band in that paper's supplementary information, although a full experimental assigned line list was not produced. In the case of the Messerle--Krauss band, recent unpublished investigations  \citep{KlaasTim} suggests that the observed lines of the Messerle--Krauss band are actually part of the Deslandres--d'Azambuja (\Cstate{} $-$ \Astate{}) band and that the true \Cprimestate{} is much higher in energy, as consistently predicted by \abinitio{} theory. 

The full \Marvel{} input file with formatted assigned transitions includes \notrans{} transitions with 6 quantum numbers following the formatting of the original \ce{C2} \Marvel{} transitions file. It is provided as supplementary information with the latest update available online on \Marvel online, \edited{\url{http://kkrk.chem.elte.hu/marvelonline/}}.

\section{Updated \Marvel{} data}

\subsection{Spectroscopic Network} 

The experimental spectroscopic network of $^{12}$C$_2$ built from assigned 
transitions has one main component
with \noenergy{} energy levels, incorporating 2061 singlet, 
4910 triplet, and 76 quintet states. 
These energy levels span \noelec{} electronic states and \novibronic{} vibronic levels. 
There are 203 other spectroscopic networks, none of which have more than 12 energy levels. 
Therefore, they are not considered further in this paper. 

\begin{figure}
    \centering
      \includegraphics[width=0.48\textwidth]{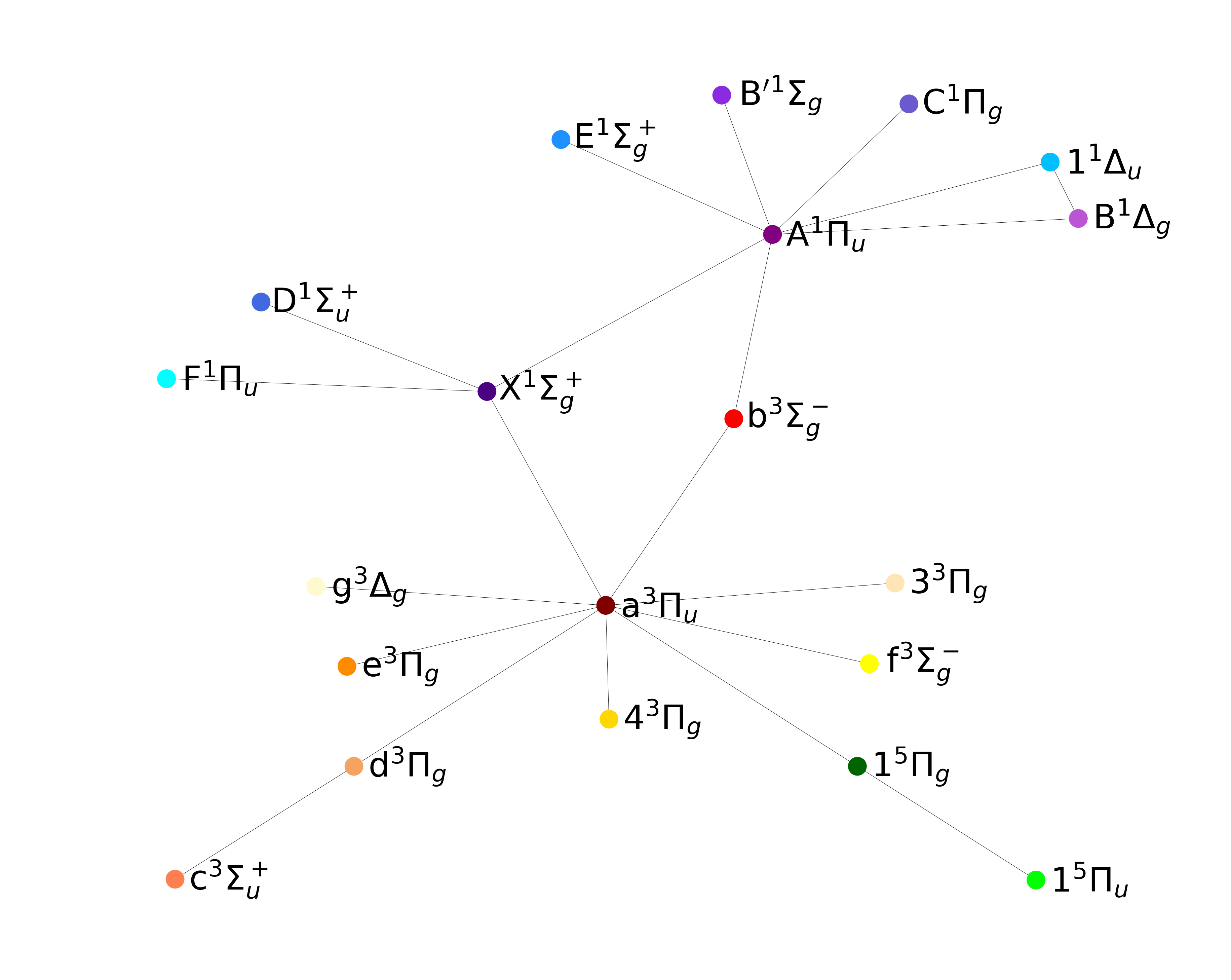}
      \includegraphics[width=0.48\textwidth]{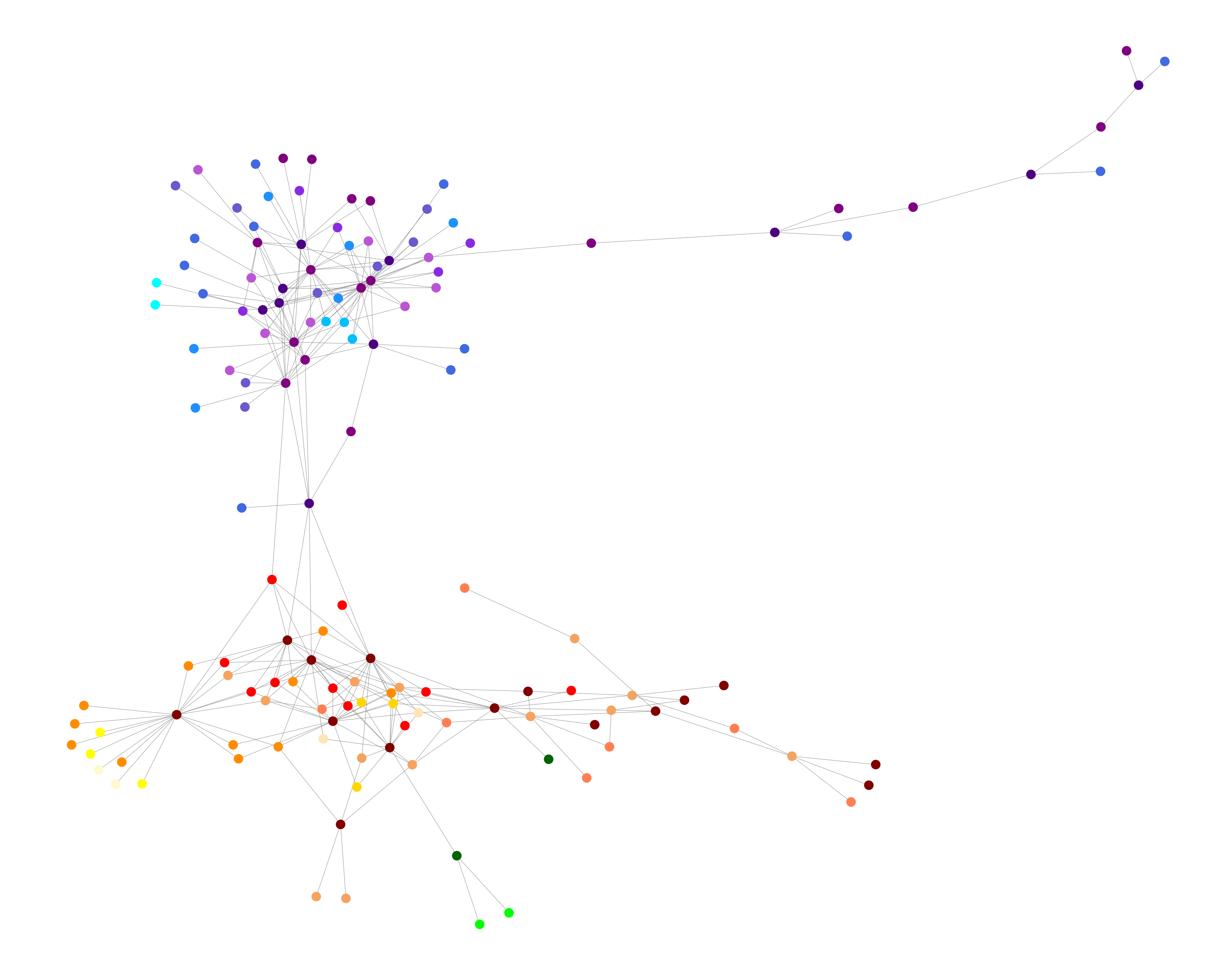}
      \includegraphics[width=0.48\textwidth]{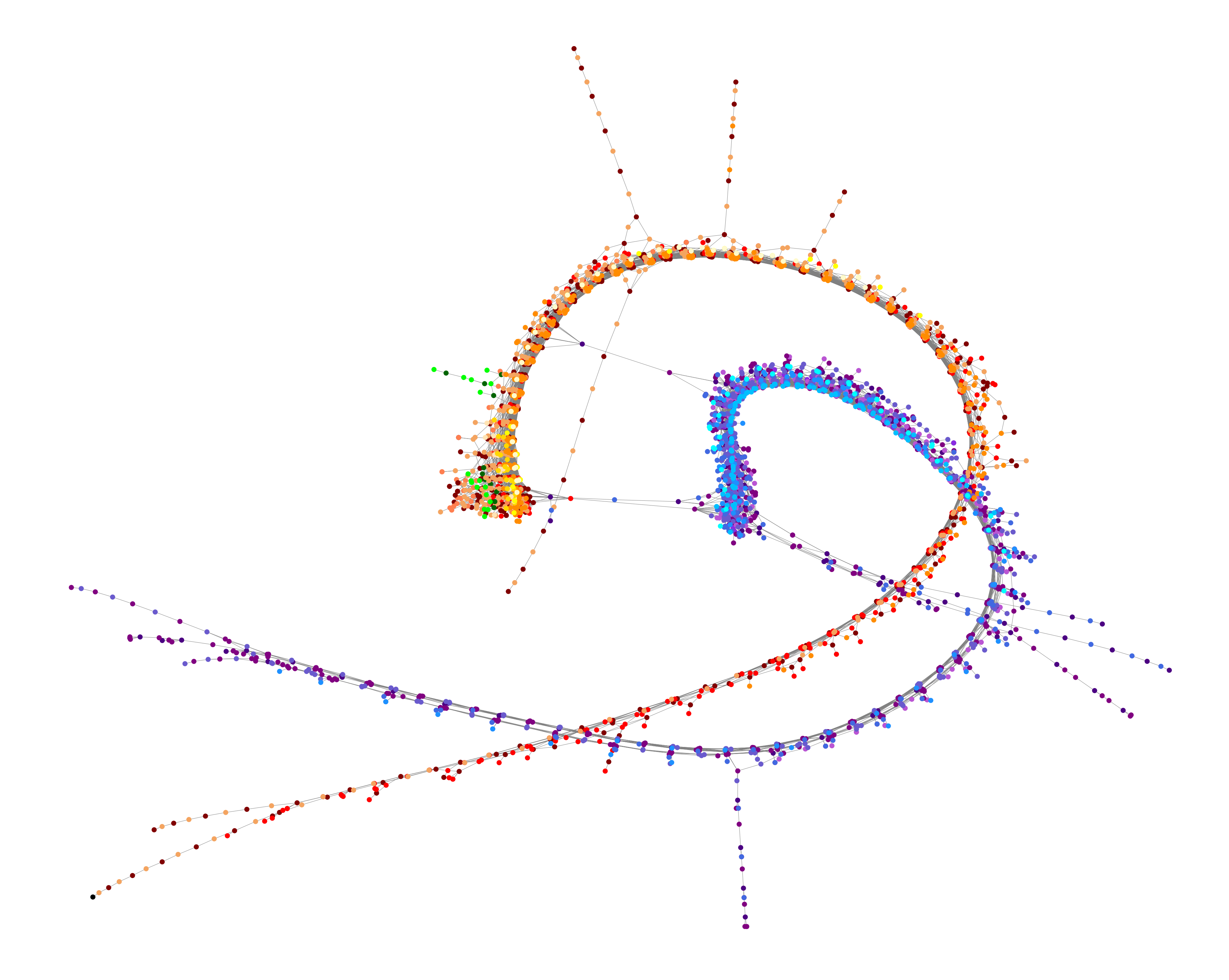}
    \caption{Spectroscopic networks of $^{12}$C$_2$ produced using the 
    \Marvel{} input and output data. Each electronic state is given a colour which is labelled in top subfigure (network of electronic states) and repeated in middle subfigure (network of vibronic states) and bottom subfigure (network of rovibronic states). The bluer colours are singlets, the redder colours are triplets and the green colours are quintets. }
    \label{fig:networks}
\end{figure}

\Cref{fig:networks} visually demonstrates the connectivity of the 
experimental spectroscopic network of $^{12}$C$_2$. 
In all of the subfigures, the nodes are the energy levels, 
and the edges are the transitions between them. 
The three subfigures show the SN at the resolution of 
electronic  (top), vibronic  (middle) and 
rovibronic (bottom) states, with the same colour scheme 
used for all figures. 

The top sub-figure of \Cref{fig:networks} shows that the singlet and triplet manifolds are largely independent, 
as expected, with two different spin-forbidden bands connecting the manifolds. 
The quintet manifold is connected to the whole spectroscopic network \emph{via} 
transitions between the $1 {}^5\Pi_g$ and \astate{} states.
Most of the triplet-bands studies have  \astate{} as the lower-energy state, 
whereas for the singlet bands the \Astate{} state is more connected 
than the \Xstate{} state; this follows from the symmetries of the various states 
with $\Pi$ states being most connected. 

The separation of the singlet and triplet manifolds becomes more pronounced 
in the middle sub-figure of \Cref{fig:networks}. 
This figure also clarifies that vibronic states are strongly interconnected in $C_2$, \emph{i.e.}, 
a given upper state can decay to many different vibrational ground states. These Franck--Condon transitions are numerous due to significant variation 
in the bond lengths of the different electronic states. 

The bottom sub-figure of \Cref{fig:networks} shows clearly the sparsity of the intercombination bands
in contrast to the other observed transitions, 
yet highlights their importance in connecting the singlet, triplet,
and quintet manifolds. 
The spiral networks of transitions in both the singlet and triplet manifolds are
multicoloured as they are between a variety of electronic states, 
with their linear structure largely determined by angular momentum 
selection rules $\Delta J = 0, \pm 1$. 
Note that we have excluded all energy levels and transitions not in the 
main component from  this figure for clarity.

\begin{table*}[htpb!]
    \sisetup{round-mode=places,retain-explicit-plus}
    \centering
    \small
    \caption{\label{tab:Esing} 
    Summary of  experimentally-derived \Marvel{} energy levels, 
    including uncertainties and data sources, for low-lying singlet states
    of $^{12}$C$_2$. Bold indicates new data sources, italics indicates updated data sources. No is the number of energy levels in that vibronic state. See  \Cref{tab:bands}  for the citations to these sources.}
\resizebox{2\columnwidth}{!}{%
\begin{tabular}{lrcrccp{10cm}}

\toprule

    State & $v$ & $J$-range & No & \mc{1}{c}{$E$-range}  & \mc{1}{c}{Av Unc (\cm{})} & Sources \\
\midrule
\Xstate{} & 0 & $0-74$ & 38 & $0-9836$ & 0.0019 & 39Landsver, 63BaRab, \textbf{69HeLaMa}, 77ChMaMa, 97SoBlLiXu, 04ChYeWoLi, 06PeSi, 15ChKaBeTa, \textbf{19Nakajima} \\ 
\Xstate{} & 1 & $0-72$ & 37 & $1827-11\,056$ & 0.0025 & 39Landsver, 63BaRab, 77ChMaMa, 88DoNiBea, 97SoBlLiXu, 04ChYeWoLi, 06PeSi, 15ChKaBeTa, \textbf{19Nakajima} \\ 
\Xstate{} & 2 & $0-58$ & 30 & $3627-8105$ & 0.0010 & 39Landsver, 63BaRab, 77ChMaMa, 88DaAbPh, 88DoNiBea, 97SoBlLiXu, 04ChYeWoLi, 15ChKaBeTa, \textbf{18KrWeFr} \\ 
\Xstate{} & 3 & $0-46$ & 24 & $5397-9158$ & 0.0021 & 39Landsver, 63BaRab, 88DaAbPh, 88DoNiBea, 97SoBlLiXu, 04ChYeWoLi, 15ChKaBeTa, \textbf{18KrWeFr} \\ 
\Xstate{} & 4 & $0-40$ & 21 & $7136-9964$ & 0.0014 & 88DaAbPh, 88DoNiBea, 95BlLiSo, 04ChYeWoLi, 13NaEn, 15ChKaBeTa, \textbf{18KrWeFr}, \textbf{19Nakajima} \\ 
\Xstate{} & 5 & $0-30$ & 16 & $8844-10\,433$ & 0.0015 & 88DoNiBea, 13NaEn, 15ChKaBeTa, \textbf{18KrWeFr}, \textbf{19Nakajima} \\ 
\Xstate{} & 6 & $0-26$ & 13 & $10\,518-11\,704$ & 0.0044 & 15ChKaBeTa, \textbf{18KrWeFr} \\ 
\Xstate{} & 7 & $0-10$ & 6 & $12\,155-12\,339$ & 0.0033 & 13NaEn, \textbf{18KrWeFr} \\ 
\Xstate{} & 8 & $0-12$ & 7 & $13\,751-14\,008$ & 0.0051 & 13NaEn, \textbf{18KrWeFr} \\ 
\Xstate{} & 9 & $0-12$ & 7 & $15\,303-15\,556$ & 0.0052 & 13NaEn, \textbf{18KrWeFr} \\ 
\vspace{-0.5em} \\
\Astate{} & 0 & $1-72$ & 72 & $8272-16\,541$ & 0.0084 &  \textbf{30DiLo}, 51Freymark, 63BaRab, 77ChMaMa, 88DoNiBeb, \textbf{88GoCo}, 88DaAbPh, 88DoNiBea, \emph{16ChKaBeTa},  15ChKaBeTa \\ 
\Astate{} & 1 & $1-79$ & 77 & $9856-19\,657$ & 0.0140 &  \textbf{30DiLo}, \textbf{40HeSu}, 51Freymark,77ChMaMa, 88DoNiBeb, \textbf{88GoCo}, 88DaAbPh, 88DoNiBea, 06PeSi, 15ChKaBeTa, \emph{16ChKaBeTa}    \\ 
\Astate{} & 2 & $1-70$ & 70 & $11\,415-19\,073$ & 0.0023 & \textbf{30DiLo}, \textbf{40HeSu}, 51Freymark, 77ChMaMa, 88DaAbPh, 88DoNiBeb, \textbf{88GoCo}, 97SoBlLiXu, 04ChYeWoLi, 06PeSi, 15ChKaBeTa,  \emph{16ChKaBeTa} \\ 
\Astate{} & 3 & $1-75$ & 75 & $12\,951-21\,607$ & 0.0047 & \textbf{30DiLo}, \textbf{40HeSu}, 51Freymark, 63BaRab, 77ChMaMa, 88DoNiBea,  \textbf{88GoCo}, 97SoBlLiXu, 04ChYeWoLi, 15ChKaBeTa, \emph{16ChKaBeTa}  \\ 
\Astate{} & 4 & $1-74$ & 74 & $14\,462-22\,800$ & 0.0059 & \textbf{40HeSu}, 51Freymark, 63BaRab, 77ChMaMa, 88DoNiBea,  \textbf{88GoCo}, 04ChYeWoLi, 15ChKaBeTa, \emph{16ChKaBeTa} \\ 
\Astate{} & 5 & $1-59$ & 59 & $15\,949-19\,784$ & 0.0086 & \textbf{40HeSu}, 63BaRab, 88DoNiBea, 04ChYeWoLi, 15ChKaBeTa, \emph{16ChKaBeTa} \\ 
\Astate{} & 6 & $1-47$ & 47 & $17\,411-20\,768$ & 0.0085 & \emph{50Phillips}, 63BaRab, 04ChYeWoLi, 15ChKaBeTa, \textbf{19Nakajima}\\ 
\Astate{} & 7 & $1-20$ & 20 & $18\,849-19\,469$ & 0.0034 & 04ChYeWoLi, 15ChKaBeTa \\ 
\Astate{} & 8 & $2-19$ & 18 & $20\,268-20\,816$ & 0.0047 & 04ChYeWoLi \\ 
\Astate{} & 9 & $1-31$ & 30 & $21\,650-23\,080$ & 0.0030 & 13NaEn, \textbf{19Nakajima} \\ 
\Astate{} & 10 & $1-30$ & 30 & $23\,013-24\,338$ & 0.0027 & 13NaEn, \textbf{19Nakajima} \\ 
\Astate{} & 11 & $1-23$ & 22 & $24\,352-25\,128$ & 0.0024 & 13NaEn, \textbf{19Nakajima} \\ 
\Astate{} & 12 & $1-7$ & 6 & $25\,665-25\,740$ & 0.0018 & 13NaEn \\ 
\Astate{} & 13 & $1-7$ & 6 & $26\,953-27\,027$ & 0.0018 & 13NaEn \\ 
\Astate{} & 14 & $1-5$ & 5 & $28\,215-28\,253$ & 0.0010 & 13NaEn \\
\Astate{} & 15 & $1-3$ & 3 & $29\,452-29\,465$ & 0.0011 & 13NaEn \\
\Astate{} & 16 & $1-4$ & 4 & $30\,662-30\,686$ & 0.0019 & 13NaEn \\

\vspace{-0.5em} \\
\Bstate{} & 0 & $2-47$ & 46 & $11\,868-15\,110$ & 0.0014 & 88DoNiBeb, \emph{16ChKaBeTa} \\ 
\Bstate{} & 1 & $2-50$ & 48 & $13\,252-16\,870$ & 0.0016 & 88DoNiBeb, \textbf{89GoCo}, \emph{16ChKaBeTa} \\ 
\Bstate{} & 2 & $2-42$ & 40 & $14\,614-17\,152$ & 0.0014 & 88DoNiBeb, \emph{16ChKaBeTa} \\ 
\Bstate{} & 3 & $2-39$ & 37 & $15\,953-18\,120$ & 0.0012 & 88DoNiBeb, \emph{16ChKaBeTa} \\ 
\Bstate{} & 4 & $2-36$ & 35 & $17\,269-19\,097$ & 0.0013 & 88DoNiBeb, \emph{16ChKaBeTa} \\ 
\Bstate{} & 5 & $2-33$ & 31 & $18\,562-20\,084$ & 0.0016 & 88DoNiBeb, \emph{16ChKaBeTa} \\ 
\Bstate{} & 6 & $2-36$ & 32 & $19\,833-21\,616$ & 0.0021 & \emph{16ChKaBeTa} \\ 
\Bstate{} & 7 & $2-34$ & 33 & $21\,081-22\,654$ & 0.0021 & \emph{16ChKaBeTa} \\ 
\Bstate{} & 8 & $3-24$ & 20 & $22\,314-23\,088$ & 0.0032 & \emph{16ChKaBeTa} \\ 
\vspace{-0.5em} \\
\Bprimestate{}& 0 & $0-32$ & 17 & $15\,197-16\,747$ & 0.0008 & 88DoNiBeb \\ 
\Bprimestate{}& 1 & $0-30$ & 16 & $16\,617-17\,974$ & 0.0008 & 88DoNiBeb \\ 
\Bprimestate{}& 2 & $2-28$ & 14 & $18\,045-19\,215$ & 0.0012 & 88DoNiBeb \\ 
\Bprimestate{}& 3 & $0-20$ & 11 & $19\,458-20\,064$ & 0.0011 & 88DoNiBeb \\
\Bprimestate{}& 4 & $0-30$ & 16 & $20\,878-22\,208$ & 0.0016 & \emph{16ChKaBeTa}  \\
\bottomrule
\end{tabular}}
\end{table*}

\begin{table*}[htpb!]
    \sisetup{round-mode=places,retain-explicit-plus}
    \centering
    \caption{\label{tab:Etrip} Summary of  experimentally-derived \Marvel{} energy levels, including uncertainties and data sources, for low-lying triplet states of  $^{12}$C$_2$. Bold indicates new data sources, italics indicates updated data sources. No is the number of energy levels in that vibronic state. See  \Cref{tab:bands} for the citations to these sources.}
\resizebox{2\columnwidth}{!}{%
\begin{tabular}{lrcrccp{10cm}}
\toprule


    State & $v$ & $J$-range & No & \mc{1}{c}{$E$-range}  & \mc{1}{c}{Av Unc (\cm{})} & Sources \\
 \midrule   
 \astate{} & 0 & $0-80$ & 226 & $604-11\,117$ & 0.0020 &   \textbf{69HeLaMa}, 79AmChMa, 83Amiot, 85RoWaMiVe, 85SuSaHi, 88DaAbSa, 94PrBe, 98BrHaKoCr, 99LlEw, 06PeSi, 07TaHiAm, 15ChKaBeTa, \textbf{17WeKrNaBa} \\ 
 \astate{} & 1 & $0-70$ & 192 & $2222-9844$ & 0.0020 &  \emph{49Phillips}, 79AmChMa, 85CuSa, 85RoWaMiVe, 85YaCuMeCa, 94PrBe, 03KaYaGuYu, 06PeSi, 07TaHiAm, 15ChKaBeTa \\ 
 \astate{} & 2 & $0-60$ & 173 & $3816-9563$ & 0.0023 & \emph{48Phillips}, \emph{49Phillips}, 79AmChMa, 85RoWaMiVe, 85YaCuMeCa, 94PrBe, 06PeSi, 07TaHiAm, 15ChKaBeTa, 15KrBaTrNa, \textbf{17KrWeBa}, \textbf{17WeKrNaBa} \\ 
 \astate{} & 3 & $0-58$ & 163 & $5388-10\,529$ & 0.0059 & \emph{37FoHe}, \emph{48Phillips}, \emph{49Phillips}, 79AmChMa, 85RoWaMiVe, 85YaCuMeCa, 94PrBe, 07TaHiAm, 10BoKnGe, 15ChKaBeTa, 15KrBaTrNa, 16KrBaWeNa, \textbf{17KrWeBa}, \textbf{17WeKrNaBa}\\ 
 \astate{} & 4 & $0-42$ & 122 & $6936-9607$ & 0.0061 & \emph{37FoHe}, \emph{48Phillips}, 79AmChMa, \emph{86HaWi}, 94PrBe, 07TaHiAm, 11BoSyKnGe, 15ChKaBeTa, \textbf{17KrWeBa}, \textbf{17WeKrNaBa}  \\ 
 \astate{} & 5 & $0-42$ & 121 & $8460-11\,102$ & 0.0233 & \emph{37FoHe}, \emph{48Phillips}, \emph{86HaWi}, 07TaHiAm, 11BoSyKnGe, 15ChKaBeTa, \textbf{17KrWeBa} \\ 
 \astate{} & 6 & $0-36$ & 104 & $9962-11\,991$ & 0.0224 & \emph{37FoHe}, \emph{48Phillips}, 07TaHiAm, 15ChKaBeTa, \textbf{17BoViBeKn}, \textbf{17KrWeBa} \\ 
 \astate{} & 7 & $0-26$ & 74 & $11\,440-12\,585$ & 0.0076 & 02TaAm, 07TaHiAm, 14NaEn \\ 
 \astate{} & 8 & $0-34$ & 100 & $12\,894-15\,210$ & 0.0048 & \emph{48Phillips}, \emph{86HaWi}, 02TaAm, 07TaHiAm, 13BoSyKnGe, 13NaEn, 13YeChWa  \\ 
 \astate{} & 9 & $0-35$ & 97 & $14\,326-16\,402$ & 0.0071 & \emph{48Phillips}, 02TaAm, 07TaHiAm, 13NaEn, 13YeChWa \\ 
 \astate{} & 10 & $0-5$ & 13 & $15\,734-15\,796$ & 0.0022 & 13NaEn \\ 
 \astate{} & 11 & $0-7$ & 15 & $17\,118-17\,182$ & 0.0008 & 13NaEn \\ 
 \astate{} & 12 & $2-5$ & 6 & $18\,479-18\,509$ & 0.0011 & 13NaEn \\ 
 \astate{} & 13 & $1-6$ & 9 & $19\,817-19\,862$ & 0.0016 & 13NaEn \\ 
  \astate{} & 14 & $2-5$ & 4 & $21\,132-21\,161$ & 0.0017 & 13NaEn \\ 

\vspace{-0.5em} \\
\bstate{} & 0 & $0-75$ & 106 & $6250-14\,542$ & 0.0015 & 79AmChMa, 85RoWaMiVe, 85YaCuMeCa, 88DaAbSa, 15ChKaBeTa \\ 
\bstate{} & 1 & $0-70$ & 102 & $7698-14\,671$ & 0.0020 & 79AmChMa, 85RoWaMiVe, 85YaCuMeCa, 15ChKaBeTa \\ 

\bstate{} & 2 & $0-70$ & 104 & $9124-16\,018$ & 0.0018 & 79AmChMa, 85RoWaMiVe, 85YaCuMeCa, 15ChKaBeTa \\ 
\bstate{} & 3 & $0-70$ & 99 & $10\,528-17\,342$ & 0.0024 & 79AmChMa, 85RoWaMiVe, 06PeSi, 15ChKaBeTa  \\ 
\bstate{} & 4 & $2-60$ & 87 & $11\,910-16\,874$ & 0.0026 & 79AmChMa,  85RoWaMiVe, 06PeSi, 15ChKaBeTa \\ 
\bstate{} & 5 & $1-58$ & 84 & $13\,270-17\,856$ & 0.0034 & 79AmChMa, 85RoWaMiVe, 06PeSi, 15ChKaBeTa \\ 
\bstate{} & 6 & $1-58$ & 80 & $14\,608-19\,140$ & 0.0047 & 79AmChMa, 15ChKaBeTa \\ 
\bstate{} & 7 & $2-40$ & 56 & $15\,924-18\,053$ & 0.0047 & 79AmChMa, 15ChKaBeTa \\ 
\bstate{} & 8 & $2-30$ & 42 & $17\,219-18\,394$ & 0.0070 & 15ChKaBeTa \\ 
\bstate{} & 19 & $17-22$ & 8 & $30\,416-30\,595$ & 0.0100 & 11BoSyKnGe \\ 
\vspace{-0.5em} \\

 \cstate{} & 0 & $0-19$ & 25 & $9280-10\,006$ & 0.0018 & 07JoNaRe, 13ChYeWa, 13NaEn \\ 
 \cstate{} & 1 & $0-24$ & 33 & $11\,312-12\,365$ & 0.0018 & 07JoNaRe, 13ChYeWa, 13NaEn, 14NaEn \\ 
 \cstate{} & 2 & $0-10$ & 14 & $13\,315-13\,482$ & 0.0010 & 07JoNaRe, 13NaEn \\ 
 \cstate{} & 3 & $0-9$ & 14 & $15\,288-15\,452$ & 0.0006 & 07JoNaRe, 13NaEn \\ 
 \cstate{} & 5 & $0-5$ & 8 & $19\,139-19\,191$ & 0.0008 & 13NaEn \\ 
 \cstate{} & 6 & $0-6$ & 9 & $21\,015-21\,065$ & 0.0008 & 13NaEn \\ 
 \cstate{} & 7 & $0-6$ & 8 & $22\,854-22\,904$ & 0.0008 & 13NaEn \\ 
\vspace{-0.5em} \\

 \dstate{} & 0 & $0-81$ & 224 & $19\,984-31\,555$ & 0.0054 & \emph{48Phillips}, 83Amiot, 85CuSa, 94PrBe, 99LlEw, 03KaYaGuYu, 07TaHiAm \\ 
 \dstate{} & 1 & $0-53$ & 159 & $21\,738-26\,809$ & 0.0070 & \emph{48Phillips}, 85SuSaHi, 94PrBe, 07TaHiAm, 13ChYeWa \\ 
 \dstate{} & 2 & $0-41$ & 119 & $23\,454-26\,237$ & 0.0094 & \emph{48Phillips}, 94PrBe,  07TaHiAm, 13ChYeWa, 13NaEn  \\ 
 \dstate{} & 3 & $0-39$ & 106 & $25\,130-27\,615$ & 0.0047 & 07JoNaRe, 07TaHiAm, 13NaEn, 14NaEn, \emph{48Phillips}, 94PrBe \\ 
 \dstate{} & 4 & $0-43$ & 126 & $26\,761-29\,736$ & 0.0098 & 07JoNaRe, 07TaHiAm, 10BoKnGe, 13BoSyKnGe, 13YeChWa \\ 
 \dstate{} & 5 & $0-36$ & 105 & $28\,342-30\,506$ & 0.0051 & 02TaAm, 07JoNaRe, 07TaHiAm, 13NaEn, 13YeChWa \\ 
 \dstate{} & 6 & $0-35$ & 103 & $29\,865-31\,876$ & 0.0096 & 02TaAm, 07TaHiAm, 11BoSyKnGe \\ 
 \dstate{} & 7 & $0-34$ & 93 & $31\,324-33\,291$ & 0.0080 & 02TaAm, 07JoNaRe, 07TaHiAm, 13NaEn \\ 
 \dstate{} & 8 & $0-32$ & 88 & $32\,709-34\,220$ & 0.0216 & \emph{48Phillips}, 07TaHiAm, 13NaEn, \textbf{17BoViBeKn}  \\ 
 \dstate{} & 9 & $2-33$ & 85 & $34\,013-35\,672$ & 0.0106 & \emph{48Phillips}, 07TaHiAm \\ 

 \dstate{} & 10 & $0-34$ & 94 & $35\,234-37\,143$ & 0.0035 & \emph{48Phillips}, 07TaHiAm, 13NaEn\\ 
 \dstate{} & 11 & $1-8$ & 13 & $36\,377-36\,458$ & 0.0015 & 13NaEn \\ 
 \dstate{} & 12 & $0-9$ & 19 & $37\,453-37\,553$ & 0.0008 & 13NaEn \\ 
 \bottomrule 
\end{tabular}}
\end{table*}

\begin{table*}
    \sisetup{round-mode=places,retain-explicit-plus}
    \centering
    \caption{\label{tab:Ehigh} Summary of  experimentally-derived \Marvel{} energy levels, including uncertainties and data sources, for highly-excited states of  $^{12}$C$_2$. Bold indicates new data sources, italics indicates updated data sources. No is the number of energy levels in that vibronic state. See  \Cref{tab:bands} for the citations to these sources.}
\resizebox{2\columnwidth}{!}{%
\begin{tabular}{lrcrccp{10cm}}

\toprule

    State & $v$ & $J$-range & No & \mc{1}{c}{$E$-range}  & Av Unc (\cm{}) & Sources \\
\midrule

$1~{}^5\Pi_g$ & 0 & $0-12$ & 30 & $29\,861-30\,082$ & 0.0047 & 11BoSyKnGe, 15BoMaGo, \textbf{17BoViBeKn} \\ 
\vspace{-0.5em} \\

 \Cstate{} & 0 & $1-78$ & 77 & $34\,241-44\,905$ & 0.1054 & \textbf{30DiLo} \\ 
 \Cstate{} & 1 & $1-71$ & 71 & $36\,005-44\,500$ & 0.0927 & \textbf{30DiLo} \\ 
 \Cstate{} & 2 & $3-67$ & 64 & $37\,719-45\,336$ & 0.1197 & \textbf{30DiLo} \\ 
 \Cstate{} & 3 & $1-24$ & 24 & $39\,306-40\,306$ & 0.1532 & \textbf{40HeSu} \\ 
 \Cstate{} & 4 & $1-36$ & 36 & $40\,775-42\,744$ & 0.1479 & \textbf{40HeSu} \\ 
 \Cstate{} & 5 & $1-35$ & 35 & $42\,033-44\,909$ & 0.1149 & \textbf{40HeSu} \\ 
 \Cstate{} & 6 & $1-41$ & 41 & $43\,030-45\,387$ & 0.1543 & \textbf{40HeSu} \\ 
 \Cstate{} & 7 & $1-37$ & 37 & $44\,631-46\,411$ & 0.1573 & \emph{50Phillips} \\ 

\vspace{-0.5em} \\
$1~{}^5\Pi_u$ & 0 & $0-13$ & 35 & $51\,651-51\,920$ & 0.0043 & 15BoMaGo \\ 
$1~{}^5\Pi_u$ & 1 & $1-6$ & 11 & $52\,495-52\,553$ & 0.0550 & \textbf{17BoViBeKn} \\ 
\vspace{-0.5em} \\

 \estate{} & 0 & $1-43$ & 122 & $40\,420-42\,533$ & 0.0471 & \emph{37FoHe}, \emph{49Phillips}, \emph{86HaWi}, \textbf{17WeKrNaBa} \\
 \estate{} & 1 & $1-55$ & 159 & $41\,455-44\,833$ & 0.0421 &  \emph{49Phillips}, \emph{86HaWi}, 98BrHaKoCr, \textbf{17WeKrNaBa} \\ 
 \estate{} & 2 & $1-49$ & 143 & $42\,433-4\,5059$ & 0.0304 &  \emph{49Phillips}, \textbf{17WeKrNaBa} \\ 
 \estate{} & 3 & $1-48$ & 138 & $43\,366-45\,832$ & 0.0399 &  \emph{49Phillips}, \textbf{17WeKrNaBa} \\ 
 \estate{} & 4 & $1-44$ & 123 & $44\,260-46\,291$ & 0.0461 &  \emph{49Phillips}, \textbf{17WeKrNaBa} \\ 
 \estate{} & 5 & $0-15$ & 41 & $45\,120-45\,349$ & 0.0245 & \textbf{17WeKrNaBa} \\ 
 \estate{} & 6 & $0-16$ & 40 & $45\,952-46\,209$ & 0.0272 & \textbf{17WeKrNaBa} \\ 
 \estate{} & 7 & $0-12$ & 32 & $46\,758-46\,901$ & 0.0268 & \textbf{17WeKrNaBa} \\ 
 \estate{} & 8 & $1-15$ & 40 & $47\,541-47\,787$ & 0.0280 & \textbf{17WeKrNaBa} \\ 
 \estate{} & 9 & $0-14$ & 35 & $48\,298-48\,511$ & 0.0289 & \textbf{17WeKrNaBa} \\
  \estate{} & 10 & $0-12$ & 24 & $49\,035-49\,192$ & 0.0289 & \textbf{17WeKrNaBa} \\ 

\vspace{-0.5em} \\

 \Dstate{} & 0 & $1-63$ & 32 & $43\,231-50\,457$ & 0.0837 & 39Landsver, 97SoBlLiXu \\ 
 \Dstate{} & 1 & $1-65$ & 33 & $45\,033-52\,631$ & 0.0942 & 39Landsver, 97SoBlLiXu \\ 
 \Dstate{} & 2 & $1-51$ & 26 & $46\,806-51\,483$ & 0.1091 & 39Landsver, 97SoBlLiXu \\ 
 \Dstate{} & 3 & $3-37$ & 18 & $48\,569-51\,014$ & 0.0979 & 39Landsver, 97SoBlLiXu \\ 
 \Dstate{} & 4 & $1-41$ & 20 & $50\,268-53\,245$ & 0.0306 & 95BlLiSo, \textbf{18KrWeFr} \\ 
 \Dstate{} & 5 & $1-19$ & 10 & $51\,956-52\,608$ & 0.0169 & \textbf{18KrWeFr} \\ 
 \Dstate{} & 6 & $1-11$ & 6 & $53\,616-53\,838$ & 0.0250 & \textbf{18KrWeFr} \\ 
 \Dstate{} & 7 & $1-11$ & 6 & $55\,247-55\,466$ & 0.0276 & \textbf{18KrWeFr} \\ 
 \Dstate{} & 8 & $1-11$ & 6 & $56\,849-57\,065$ & 0.0321 & \textbf{18KrWeFr} \\ 
 \Dstate{} & 9 & $1-11$ & 6 & $58\,422-58\,635$ & 0.0101 & \textbf{18KrWeFr} \\ 
 
 \Dstate{} & 10 & $1-11$ & 6 & $59\,965-60\,175$ & 0.0101 & \textbf{18KrWeFr} \\ 
 \Dstate{} & 11 & $1-11$ & 6 & $61\,478-61\,686$ & 0.0091 & \textbf{18KrWeFr} \\ 
\vspace{-0.5em} \\

 \threestate{} & 3 & $1-8$ & 23 & $5\,0681-50\,781$ & 0.0743 & \textbf{17KrWeBa} \\ 
 \threestate{} & 8 & $1-9$ & 24 & $54\,566-54\,643$ & 0.0918 & \textbf{17KrWeBa} \\ 
\vspace{-0.5em} \\
 \fourstate{} & 0 & $1-9$ & 26 & $52\,106-52\,236$ & 0.0696 & 15KrBaTrNa, 16KrBaWeNa, \textbf{17KrWeBa} \\ 
 \fourstate{} & 1 & $1-8$ & 20 & $53\,375-53\,460$ & 0.0846 & 15KrBaTrNa, \textbf{17KrWeBa} \\ 
 \fourstate{} & 2 & $1-7$ & 20 & $54\,699-54\,779$ & 0.0836 & \textbf{17KrWeBa} \\ 
\vspace{-0.5em} \\

 \Estate{}& 0 & $0-66$ & 32 & $54\,937-62\,593$ & 0.0610 & 51Freymark, 97SoBlLiXu \\ 
 \Estate{}& 1 & $0-68$ & 28 & $56\,529-64\,443$ & 0.0655 & 51Freymark, 97SoBlLiXu \\ 
 \Estate{}& 2 & $4-32$ & 15 & $58\,077-59\,816$ & 0.0863 & 51Freymark \\ 
 \Estate{}& 3 & $6-36$ & 13 & $59\,550-61\,658$ & 0.0910 & 51Freymark \\ 
 \Estate{}& 4 & $2-46$ & 15 & $60\,855-64\,263$ & 0.0902 & 51Freymark \\ 
 \Estate{}& 5 & $10-30$ & 11 & $62\,305-63\,573$ & 0.0787 & 51Freymark \\ 
\vspace{-0.5em} \\
$1~{}^1\Delta_u$ & 0 & $1-33$ & 61 & $57\,374-58\,868$ & 0.1388 & \textbf{88GoCo}, \textbf{89GoCo} \\ 
$1~{}^1\Delta_u$ & 1 & $2-37$ & 35 & $58\,481-60\,317$ & 0.1015 & \textbf{88GoCo} \\ 
$1~{}^1\Delta_u$ & 2 & $2-31$ & 30 & $59\,546-60\,814$ & 0.1416 & \textbf{88GoCo} \\ 
\vspace{-0.5em} \\

\fstate & 0 & $3-31$ & 15 & $70\,819-72\,208$ & 0.0843 & \textbf{69HeLaMa} \\ 
\fstate & 1 & $5-31$ & 14 & $72\,176-73\,518$ & 0.0819 & \textbf{69HeLaMa} \\ 
\fstate & 2 & $3-23$ & 11 & $73\,452-74\,198$ & 0.1053 & \textbf{69HeLaMa} \\ 
\vspace{-0.5em} \\
\gstate{} & 0 & $3-32$ & 85 & $72\,268-73\,762$ & 0.0789 & \textbf{69HeLaMa} \\ 
\gstate{} & 1 & $3-31$ & 80 & $73\,741-75\,203$ & 0.0833 & \textbf{69HeLaMa} \\ 
\vspace{-0.5em} \\
\Fstate & 0 & $10-38$ & 29 & $74\,713-76\,944$ & 0.0994 & \textbf{69HeLaMa} \\ 
\Fstate & 1 & $2-34$ & 29 & $76\,100-78\,003$ & 0.1099 & \textbf{69HeLaMa} \\ 
\bottomrule
\end{tabular}}
\end{table*}

\begin{figure*}
    \centering
    \includegraphics[width = 0.9\textwidth]{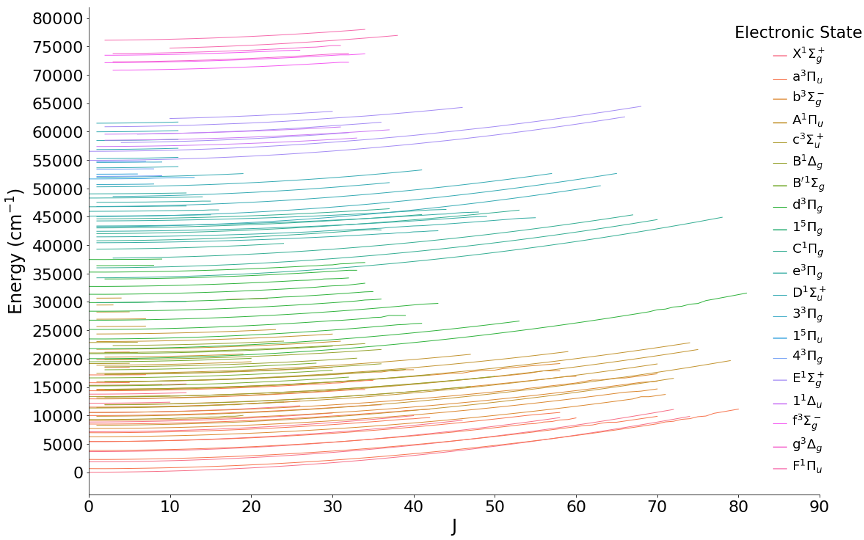}
    \caption{Summary of the energy levels from the \Marvel{} procedure. Each line is a unique spin-vibronic state, with each electronic state a unique colour.}
    \label{fig:J-energy}
\end{figure*}

\subsection{Updated \Marvel{} Energy Levels}

\Cref{fig:J-energy} gives a summary of all \noenergy{} empirical energy levels
determined in this study, 
with each line representing energy levels of a single vibronic state as 
a function of the total angular momentum quantum number $J$. 
These curves are clearly quadratic and smooth, 
suggesting that there are no major issues with the empirical energy levels. 

In \Cref{tab:Esing,tab:Etrip,tab:Ehigh}, we describe the updated \Marvel{} dataset for each vibronic level for low-lying singlet states, low-lying triplet states and higher energy states respectively in terms of
(a) the range of total angular momentum quantum numbers $J$ and energies included; 
(b) the total number of quantum states included;
(c) the average uncertainty of the derived energies; and
(d) the contributing data sources. 

\begin{figure*}
    \centering
    \includegraphics[width =1\textwidth]{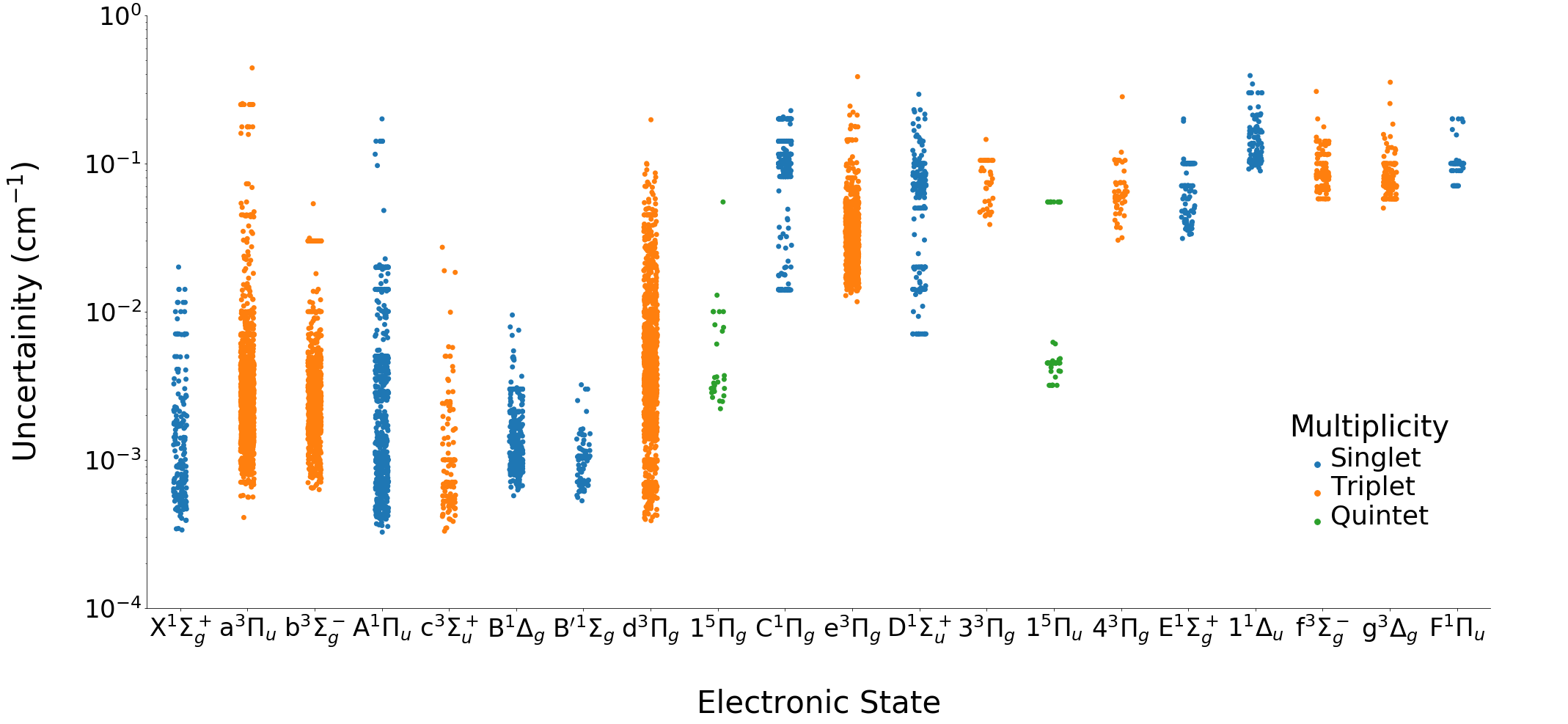}
    \caption{The distribution of uncertainties of the empirical energy levels generated for each electronic energy level.}
    \label{fig:state-unc}
\end{figure*}

\begin{figure*}
    \centering
    \includegraphics[width=\textwidth]{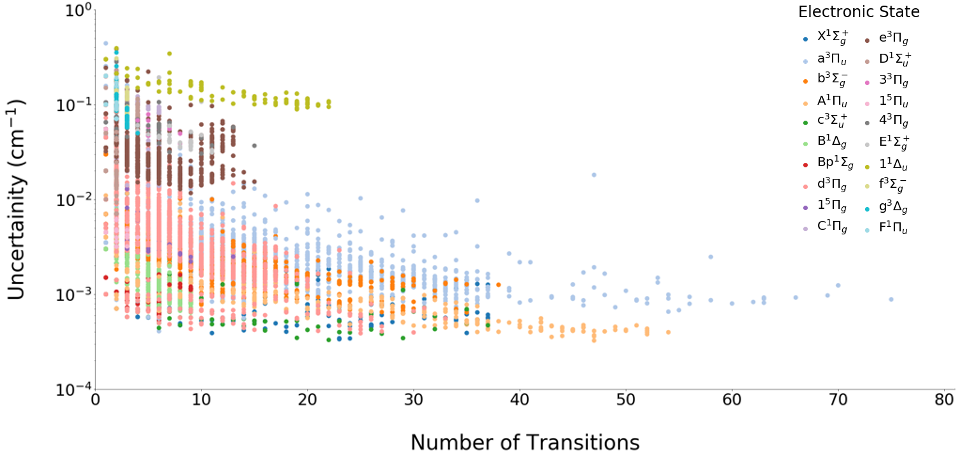}
    \caption{The relationship between the final uncertainty of the empirical energy levels and the number of transitions that contributed to the derived energy level. }
    \label{fig:trans-unc}
\end{figure*}

Low-lying states, particularly the \Xstate ($v=0-5$), \astate{} ($v=0-6$), 
and \Astate ($v=0-5$) ones, are very well characterised to high $J$ values
with data from multiple sources leading to low uncertainties usually averaging 
less than 0.002 \cm{}. 
Higher-lying states generally are characterised by far fewer sources and 
usually exhibit a much more limited range in $J$. 
States with energies lower than approximately 30\,000 \cm{} generally retain 
low median uncertainties on the order of 0.002 \cm{}, 
while the higher-lying states detailed in \Cref{tab:Ehigh} generally have 
median uncertainties of $0.01-0.1$ \cm{}, reflecting the lower accuracy of 
ultraviolet-spectroscopy measurements.

\subsection{Uncertainties}

\Cref{fig:state-unc} shows the uncertainties of the energy levels in each electronic state, ordered from left to right in order of increasing energy. 
The eight low-lying electronic states, \Xstate{}, \astate{}, \bstate{}, \Astate{}, \cstate{}, \Bstate{}, \Bprimestate{}, and \dstate{}, 
have uncertainties that range from less than 0.001 \cm{} to a small number of 
transitions with uncertainties greater than 0.01 \cm{}.
These low uncertainties are definitely suitable for the current needs of
high-resolution astronomical spectroscopy.
As an example, we note recent trends in using cross-correlation of template 
and measured spectra to extract very small data signals, e.g. for non-dominant isotopologues \citep{19MoSn.highres}. 
The quintet states have uncertainties around 0.002 \cm{}. 
In contrast, higher singlet and triplet electronic states have a higher uncertainty, generally $0.01-0.1$ \cm{}, with a smaller spread.
These higher uncertainties can be attributed to the lower resolution of the 
ultraviolet-spectroscopy experiments needed to characterise these high-lying states 
along with the smaller number of experimental data sources. 

It is useful to consider how the source uncertainties of the transitions 
(illustrated in \Cref{fig:source-unc}) propagate to the uncertainties 
of the energy levels (illustrated in \Cref{fig:state-unc}). 
Overall, the uncertainty in the energy levels seems to be approximately an 
order of magnitude lower than the uncertainty of the transitions.  
This relationship is explored further in \Cref{fig:trans-unc}, 
which plots the uncertainty of an energy level (in logarithmic scale) 
as a function of the number of transitions used in its determination. 
As expected, in general, as the number of transitions increases
the uncertainty of the energy level decreases.



\subsection{Differences between old and new \Marvel{} energy levels}
\subsubsection{New levels} 
Considering only the main component,
the 2020 \Marvel{} compilation of $^{12}$C$_2$ spectroscopic data
added 1524 rovibronic states (765 singlets, 747 triplets, and 12 quintets) 
to the 2016 compilation and removed 147 states (146 triplets and 1 quintet) 
due primarily to reprocessing of the 07TaHiAm data to remove the predicted unmeasured  transitions previously incorrectly included. 
55 of the removed states are still present in the 2020 compilation 
as orphan energy levels (\emph{i.e.}, outside the main spectroscopic network), 
indicating that the connections between these energy levels and 
the main component were removed by the reprocessing.  

The new energy levels span 18 of the total of 20 electronic states in the 
2020 \MARVEL{} $^{12}$C$_2$ spectroscopic data compilation,
and 79 of the \novibronic{} vibronic states. 
Six of these electronic states (\Cstate{}, \Fstate{}, \fstate{}, \gstate{}, \Onedeltastate{}, \threestate{}) and 44 of these vibronic states 
are entirely new to this 2020 update.
Increases in coverage were also notable for the \estate{} state
(increase of 333 levels across 11 vibrational states) and the \Astate{} state 
(increase of 106 levels across 12 vibrational states).

\begin{table}[]
    \centering
    \caption{Differences between old and new \Marvel{} compilations by electronic state, quantified by MAD (mean absolute deviation) and Max (maximum deviation). The "No" column specifies the number of states common to the new and old compilations.}
    \label{tab:differences}
    \begin{tabular}{lrcHccH}
\toprule
& & \mc{3}{c}{2020 -- 2016 Energies} \\
\cmidrule(r){3-5}
State & No &  MAD &   & Max & 2016 Av unc & \\
\midrule
\Xstate & 174 & 0.0034 &  & 0.0565 & 0.0006 \\
\Astate & 512 & 0.0041 & 0.0072 & 0.0566 & 0.0022 \\
\Bstate & 322 & 0.0013 & 0.0042 & 0.0565 & 0.0022 \\
\Bprimestate & 58 & 0.0006 & 0.0008 & 0.0019 & 0.0008 \\
\Dstate  & 117 & 0.0888 & 0.1388 & 0.5387 & 0.1643\\
\Estate & 113 & 0.0140 & 0.0556 & 0.4942 & 0.1498  \\
\vspace{-0.9em} \\

\astate{} & 1381 & 0.0386 & 0.2290 & 3.0007 & 0.0017 \\
\bstate{} & 762 & 0.0079 & 0.0125 & 0.0829 & 0.0016 \\
\cstate{} & 111 & 0.0038 & 0.0045 & 0.0152 & 0.0015\\
\dstate{} & 1314 & 0.0847 & 0.5129 & 7.8451 & 0.0023 \\
\estate{} & 563 & 0.0498 & 0.1234 & 1.4348 & 0.0659 \\
\fourstate{} & 32 & 0.4383 & 0.6531 & 1.4070 & 0.0863 \\

\vspace{-0.9em} \\
1 ${}^5\Pi_g$ & 29 & 0.0067 & 0.0085 & 0.0252 & 0.0116\\
1 ${}^5\Pi_u$ & 35 & 0.0065 & 0.0085 & 0.0252 & 0.0171 \\
\bottomrule
    \end{tabular}
\end{table}

\subsubsection{Changes in energy in previously included levels}
The old (2016) and new (2020) \Marvel{} rovibronic energy levels for $^{12}$C$_2$
are different, though these differences are usually small. 
This difference is quantified in \Cref{tab:differences}, which presents the average change in the energy levels of a given electronic state for quantum states that are in both the 2016 and this 2020 \Marvel{} compilations. 
The same data at vibronic resolution is provided in the Supplementary Information. 

Usually, the two \Marvel{} compilations predict energies that on average agree 
to about 0.005 \cm{} for low-lying states, with maximum deviations around 0.06 \cm{}.
For higher electronic states, the differences in empirical energies between the 
two compilations is higher, up to an average of 0.44 \cm{} for the \fourstate{} 
state following the addition of significant new data in this 2020 \Marvel{} update.  
The main outliers to this trend are the \astate{}, \dstate{}, and \estate{} states,
    which all have quite sizeable modifications from 2016 to 2020. 
    These changes can be traced primarily to a reprocessing of the 49Phillips
    and 07TaHiAm data, which corrected earlier errors. 
    We carefully examined the states with large deviations 
    and found that the energies along a vibronic band were far smoother 
    and more reasonable in the 2020 update than the original 2016 compilation, 
    indicating that these modifications improved the overall data compilation.

An integral part of the \Marvel{} process is to provide predictions for 
the accuracy of its empirical energy levels. 
Our data here shows that the 2016 prediction of the average uncertainties 
generally is quite close to the changes in energy observed between 
the 2016 and 2020 compilation, indicating that the original uncertainties were reasonable.  
The most notable underestimation in the original uncertainties is 
for the \Xstate{} state, which under-predicted changes by about a factor of 5. 
Other significant deviations between the 2016 prediction uncertainties 
and the 2020--2016 changes in energy can be attributed to mis-assignments 
and processing errors in the 2016 compilation.



\section{Updated C$_2$ line lists}

The supplementary information of this paper contains three updated states files for $^{12}$C$_2$, $^{12}$C$^{13}$C and $^{13}$C$_2$ called \texttt{12C2\_8states\_MARVEL-2020.states}, \texttt{12C-13C\_8states\_MARVEL-2020.states} and \texttt{13C2\_8states\_MARVEL-2020.states} respectively. 

The main modification to the original \texttt{8states} $^{12}C_2$ line list states file is that we re-\Marvel ised the energy levels by replacing existing energy levels with 4842 \Marvel ised energy levels from this 2020 \Marvel{} update; these \Marvel ised energy levels are denoted by a "m" while energy levels computed solely from the \Duo{} spectroscopic model are denoted by "d". 
Note that there has been no modifications to the underlying spectroscopic model  of this line list, \emph{i.e.}, the potential energy and coupling curves  were not refitted. 
For full details of the spectroscopic model for this linelist, one should refer to the original line-list paper  \citep{jt736}. 

Though we haven't compiled a \Marvel{} dataset for the C$_2$ isotopologues, a reasonable assumption is that the line list errors are similar between different isotopologues. Therefore, we follow past precedent \citep{jt665,jt760} in creating a pseudo-\Marvel ised states file for $^{12}$C$^{13}$C and $^{13}$C$_2$ by using: \begin{equation}E_\textrm{iso} \approx E_\textrm{iso}^\textrm{Duo} + (E_\textrm{main}^{\textrm{MARVEL}} - E_\textrm{main}^{\textrm{Duo}}), \end{equation} where $E_\textrm{iso}$ is the isotopologue energy for a given state in the final line list, $E_\textrm{iso}^\textrm{Duo}$ is the original spectroscopic model prediction using \Duo{} and $E_\textrm{main}^{\textrm{MARVEL}}$ and $E_\textrm{main}^{\textrm{Duo}}$ are the \Marvel{} and \Duo{} predicted energies of the state for the main isotopologue, in the case $^{12}$C$_2$. Energy levels modified in this way are labelled by "i" (for isotopologue pseudo-\Marvel isation, in this way clearly distinguished from \Marvel isation in the main isotopologue states file). For the case of C$_2$, nuclear spin statistics means that some microstates are present in the isotopologue states files that are not present in the main $^{12}$C$_2$ isotopologue states file, e.g. only one parity component in a $\Pi$ state is retained for each $J$ in  $^{12}$C$_2$ while both are present for  $^{13}$C$_2$ and  $^{12}$C$^{13}$C. To account for this, we used (\Marvel{} - \Duo) energy differences from a single $^{12}$C$_2$ parity state to correct both parity states in the isotopologue files. 

The minor modification to the \texttt{8states} $^{12}$C$_2$ states file is the inclusion of 71 energy levels that were predicted by 07TaHiAm data but not otherwise included. These were based on 746 transition frequencies from 07TaHiAm \citep{07TaHiAm.C2} which were predicted (not experimentally measured), with some of these inadvertently included in the 2016 compilation though none are in this update. Given that the  07TaHiAm predictions are based on band-specific fits, they are likely more accurate than the original \LLname{} energy levels. Therefore, we found the associated energy levels of these additional lines by creating an extended \Marvel{} input transitions file with these 746 additional frequencies (called \texttt{12C2\_experimentaland07TaHipredicted\_Marvel.inp} in the Supplementary Information) and extracted 71 additional energy levels (compiled in \texttt{Predicted07TaHiAm.energies} file in the Supplementary Information) which replaced the original energy levels in our updated $^{12}$C$_2$ \LLname{} states file, with a label "p" (for perturbed). 


Note that the process of \Marvel ising the \LLname{} linelist was shown to help identify  very subtle mistakes in the \Marvel{} compilation itself. 
For example, large errors in \Xstate{}, $v=2$, $J=50-54$ energies between 
the line-list prediction and \Marvel{} energies helped identify 
a digitisation error where a "8" was read as a "3" in the transition 77ChMaMa.558. These errors were corrected in the final set of \Marvel{} transitions, energy levels and line lists provided by the Supplementary Information. 

As detailed in \Cref{tab:MARVELisation}, 
this update increases the number of \Marvel ised energy levels 
within the line list from 4555 to 4916, increasing the coverage from 
10.3\% of all rovibronic levels to 11.1\%. 
The new \Marvel ised energy levels are primarily additional vibrational levels ($\nu=8-9,11-16$) and expanded rotational coverage within the \Astate{} state, 
with data on  \bstate{}, $n=19$ also being added for the first time. 
The total number of transitions that are \Marvel ised, 
\emph{i.e.}, those with frequencies which are entirely determined by \Marvel{} 
energy levels, remains relatively low, increasing from 4.3\%  to 5.2\%. 

\begin{figure}
    \centering
    \includegraphics[width=0.45\textwidth]{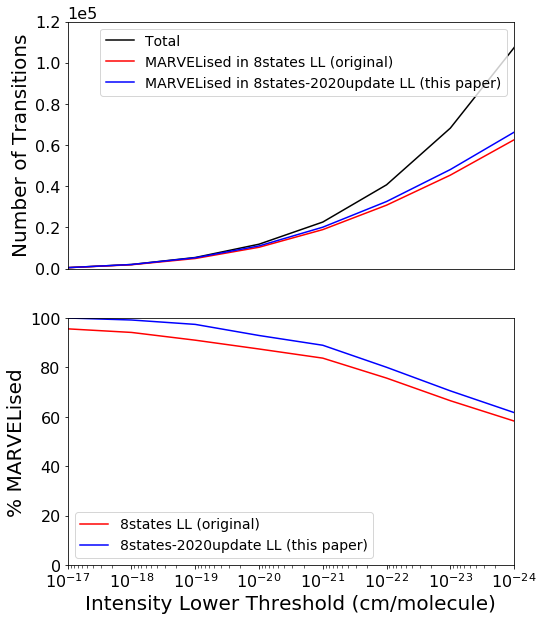}
    \caption{Proportion of strong-to-moderate intensity transitions 
    that are \Marvel ised (based on experimentally-derived \Marvel{} energy levels) 
    at 1000 K in the updated line list compared to the original \LLname{} line list. 
    Top panel considers the number of transitions (total number of transitions vs the number \Marvel ised in the original and new versions of the line list), 
    whereas the bottom panel converts this to percentage of transitions 
    that are \Marvel ised. 
    The horizontal axis of both panels is an intensity lower threshold 
    in a logarithmic scale; the vertical axis data gives data for all 
    transitions with intensities above this threshold. }
    \label{fig:perctrans}
\end{figure}

Looking at the strong to medium intensity transitions, however, 
in \Cref{fig:perctrans}, we see that the bulk of the strong transitions 
are \Marvel ised and that the 2020 update increases the percentage of \Marvel ised transitions by about 5\% when considering the percentage of transitions 
with intensities above $10^{-17}$ - $10^{-24}$ cm\,molecule$^{-1}$ at 1000 K. 
The updated line list now has experimentally-derived (\Marvel ised) transition frequencies for all strong transitions with intensities above 
10$^{-17}$ cm\,molecule$^{-1}$ (from 95.6\% in original 8states) and 
80.0\% (from 75.7\%) of all transitions with intensities above 
10$^{-22}$ cm\,molecule$^{-1}$ at 1000 K.

\begin{table}
    \caption{Summary of the overall proportion of energy levels and transitions that are \Marvel ised, i.e. based entirely on experimentally-derived \Marvel{} energy levels.}
    \label{tab:MARVELisation}
    \centering
    \begin{tabular}{rrrrrrrr}
    \toprule
    & \mc{2}{c}{\LLname{} } \\
         & Original & Update \\
         \midrule
   \mc{3}{l}{\textbf{Energy levels}} \\
        Number \Marvel ised & 4\,555 & 4\,916*\\
        Total  & 44\,189 & 44\,189 \\
        \% \Marvel ised & 10.3 & 11.1 \\
   \vspace{-0.5em} \\
      \mc{3}{l}{\textbf{All transitions (no intensity threshold)}}\\
     Number \Marvel ised & 258\,729 & 307\,076 \\
     Total Num & 6\,080\,920 & 6\,080\,920 \\
      \% \Marvel ised &  4.3 & 5.0 \\
     \bottomrule
    \end{tabular}
    \newline
\vspace{0.5em} 

    $^*$ 71 of these \Marvel ised energy levels are found by combining the 07TaHiAm predicted transition frequencies with the other \Marvel{} energy levels and running \Marvel{}.
\end{table}

Due to the very high proportion of \Marvel isation for strong- to 
moderate-strength transitions across the line list's full spectral range, 
this updated $^{12}$C$_2$ line list is suitable for cross-correlation 
high-resolution studies of $^{12}$C$_2$ in gaseous astrochemistry environments 
such as exoplanets \citep{14deBiBr.highres}.









\section{Discussion and Conclusions}
The 2016 \Marvel{} compilation has been significantly updated, with the addition of assigned transitions data 
from 8 old and 5 new experiments on $^{12}$C$_2$ to significantly extend 
the previous \Marvel{} dataset for $^{12}$C$_2$, including an extra 8072
 transitions and 1524 energy levels spanning an extra six electronic states, and extra 44 vibronic states. Data from five previously included sources have been updated and extended. The new data enabled a significant improvement to the quality of the ExoMol $^{12}$C$_2$ linelist by increasing the \Marvel{}isation-fraction of strong lines with frequencies above 10$^{-18}$ cm\,molecule$^{-1}$ from 94.2\% to 99.4\%.
This increase in high accuracy experimentally-derived (i.e., \Marvel{}) energy levels is extremely important astrophysically for very high resolution cross-correlation measurements that are now increasingly common with the new generation of ground-based ultra-large telescopes.

For laboratory spectroscopists, the existence of the \Marvel{} compilation, the \Marvel{} procedure and line lists have two main benefits: (1) a method to validate their data - for example, in this paper, use of \Marvel{} enables  calibration errors in 17KrWeBa to be identified and corrected, (2) enables their new results to be readily available to applications experts.  Inclusion of new data is best enabled by producing formatted assigned transitions, ideally adding their new data to pre-existing \Marvel{} compilations for the molecule and contacting the creators of the most recent linelist for their molecule. This strongly emphasizes the importance of authors providing their primary experimental data in the form of assigned line lists rather than just supplying compound results such as spectroscopic constants.  \Marvel{} compilations currently exist for
AlH  \citep{jt732}, BeH  \citep{jt722}, $^{12}$C$_2$  \citep{jt637}, $^{14}$NH  \citep{jt764}, 
NO \citep{jt686}, $^{16}$O$_2$  \citep{19FuHoKoSo}, $^{48}$Ti$^{16}$O  \citep{jt672},
$^{90}$Zr$^{16}$O  \citep{jt740},  isotopologues of \ce{H2O}  \citep{jt454,jt482,jt539,jt576,jt562,jt795},  H$_2$$^{32}$S  \citep{jt718}, isotopologues of \ce{H3+}  \citep{13FuSzMaFa,13FuSzFa.marvel},
isotopologues of \ce{SO2}  \citep{jt704}, \ce{C2H2}  \citep{jt705}, \ce{H2CCO}  \citep{marvel_h2cco} and \ce{NH3}  \citep{jt608,jt784}, 
with ongoing work on other molecules.

For many diatomics accurate measurement of spin-forbidden bands is very important 
for enabling the relative positioning of the spin manifolds, in the present case 
the singlet, triplet and quintet ones. 
There are 88 of these transitions known for $^{12}$C$_2$;
however, the small $T_{\rm e}$ of the lowest singlet state ($\approx 600$ \cm{})
increases the importance of this information as this means that triplet absorption
is very important for the high-resolution spectroscopy of \ce{C2}. 

All allowed electronic bands from states below 12\,000 \cm{} to a known electronic state have rotationally resolved data included in this \Marvel{} update except for the Messerle--Krauss band and the Kable--Schmidt band. 
In the former case, the initial assignment of the band is currently being questioned  \citep{KlaasTim} with the \Cprimestate{} expected by theory to be much higher in energy. In the latter case, the two electronic states  involved are well characterised by other studies and so the updated \LLname{} ExoMol line list  can be expected to provide very accurate predictions of its spectroscopy for users. 

A common trait of the higher lying states is the higher uncertainties in their energies, originating from lower resolution studies in the ultraviolet region than other visible region studies. Of particular note is the  Deslandres--d'Azambuja band system which has been  seen in flames  \citep{49HoHe}, plasma plumes  \citep{08CaDiSaRe}, laser ablation of graphite  \citep{02AcDe} and astrophysics  \citep{89GrVaBl,07BeBePi}. Yet remarkably the only modern, high resolution study of these band was made of $^{13}$C$_2$  \citep{85AnBoPe.C2}.
A high resolution study of the  Deslandres--d'Azambuja bands for $^{12}$C$_2$ is overdue.

\section*{Acknowledgments}
We thank Klaas Nauta, Tim Schmidt and Scott Kable for helpful discussions 
and feedback. 
SNY and JT's work was supported by the STFC Projects No. ST/M001334/1 and ST/R000476/1.
The work performed in Budapest received support from NKFIH (grant no. K119658) and
from the ELTE Excellence Program (1783-3/2018/FEKUTSTRAT) supported by
the Hungarian Ministry of Human Capacities (EMMI).

\section*{Data Availability Statement} 
The data underlying this article are available in the article and in its online supplementary material. These include the following file;
\begin{itemize}
    \item README\_2020\_C2\_MARVEL - Explanation of all files within the SI
    \item 12C2\_MARVEL.inp - The final \Marvel{} transitions file
    \item 12C2\_MARVEL.energies - The final \Marvel{} energies file
    \item 12C2\_experimentalplus07TaHiAmpredicted\_MARVEL.inp - The \Marvel{} input file that allows the predictions of 07TaHiAm to be incorporated into the final line list
    \item Predicted07TaHiAm.energies - The additional energies predicted from 12C2\_experimentalplus07TaHiAmpredicted\_MARVEL.inp
    \item 12C2\_\_8states\_MARVEL-2020.states, 12C13C\_\_8states\_MARVEL-2020.states and 13C2\_\_8states\_MARVEL-2020.states - The updated states file for the ExoMol 8states line list (to be used with trans file from ExoMol website)
    \item SI\_2020\_C2MARVEL.pdf - Expanded analysis tables and figures
\end{itemize}

\bibliography{All,jtj}
\bibliographystyle{mnras}




\bsp	
\label{lastpage}
\end{document}